\documentclass[a4paper]{iopart}
\usepackage{feynmp-auto}
\usepackage{color}
\usepackage{bm}
\usepackage{citesort}

\newcommand{\be}{\begin{equation}}
\newcommand{\ee}{\end{equation}}

\newcommand{\bea}{\begin{eqnarray}}
\newcommand{\eea}{\end{eqnarray}}
\newcommand{\beastar}{\begin{eqnarray}}
\newcommand{\eeastar}{\end{eqnarray}}
\newcommand{\nn}{\nonumber\\}
\newcommand{\lav}{\left\langle}
\newcommand{\rav}{\right\rangle}
\newcommand{\bsig}{\mbox{\boldmath $\sigma$}}

\newcommand{\half}{\frac{1}{2}}

\newcommand{\eq}[1]{~(\ref{#1})}
\newcommand{\eqq}[2]{~(\ref{#1},\ref{#2})}

\newcommand{\order}{{{\mathcal O}}}

\newcommand{\ie}{{i.e.}}
\newcommand{\eg}{{e.g.}}
\newcommand{\lc}{{\rm lc}}
\newcommand{\lhalf}{\mbox{$\frac{1}{2}$}}

\usepackage{url}

\newcommand{\hence}{\mbox{$\Rightarrow$}}

\newcommand{\mident}{\mbox{\boldmath $I$}}

\newcommand{\hph}{\hat\phi}
\newcommand{\p}[2]{\eta_{#1#2}}
\newcommand{\ps}[1]{\p{1}{#1}}
\newcommand{\pss}[1]{\p{2}{#1}}
\newcommand{\GG}[4]{G_{#1#2,#3#4}}
\newcommand{\GGn}[4]{G^0_{#1#2,#3#4}}
\newcommand{\Sig}[4]{\Sigma_{#1#2,#3#4}}

\newcommand{\T}{^{\rm T}}
\newcommand{\Si}{S_{\rm int}}
\newcommand{\ovr}[1]{\overline{#1}}

\begin{document}
\begin{fmffile}{diag}
\title{Path integral methods for the dynamics of stochastic and disordered systems}
\author{John A. Hertz$^{1,2}$, Yasser Roudi$^{1,3,4}$, Peter Sollich$^{5}$}

\address{$^1$ Nordita, KTH Royal Institute of Technology and Stockholm University, Stockholm, Sweden}
\address{$^2$ Department of Neuroscience and Pharmacology and Niels Bohr Institute, University of Copenhagen, Denmark}
\address{$^3$ Kavli Institute and Centre for Neural Computation Trondheim, Norway}
\address{$^4$ Institute for Advanced Study, Princeton,  USA}
\address{$^5$ King's College London, Department of Mathematics, Strand, London WC2R 2LS, UK}
% \maketitle

\begin{abstract}
We review some of the techniques used to study the dynamics of disordered systems subject to both quenched and fast (thermal) noise. Starting from the Martin-Siggia-Rose path integral formalism for a single variable stochastic dynamics, we provide a pedagogical survey of the perturbative, i.e.\ diagrammatic, approach to dynamics and how this formalism can be used for studying soft spin models. We review the supersymmetric formulation of the Langevin dynamics of these models and discuss the physical implications of the supersymmetry. We also describe the key steps involved in studying the disorder-averaged dynamics. Finally, we discuss the path integral approach for the case of hard Ising spins and review some recent developments in the dynamics of such kinetic Ising models. 
\end{abstract}

\section{Introduction}

Studying the statistical properties of variables or classical fields subject to stochastic forces has a long history in physics. A key tool in this effort is the Langevin equation \cite{langevin1908theorie}. Originally introduced for describing the Brownian motion of a particle in a fluid, over the years, this has also been used for studying the dynamics of a variety of other systems including the standard models of critical phenomena, e.g.\ those described by the Ginzburg-Landau Hamiltonian \cite{ma2000modern} and, later on, the relaxation of spin glass models with soft spins \cite{de1978dynamics,sompolinsky1981dynamic}. 

In order to study the dynamical version of the Ginzburg-Landau model in the presence of thermal noise one has to appeal to a perturbative analysis. A useful pedagogical description of the early methods used in such a perturbative treatment can be found in the seminal book by Ma \cite{ma2000modern} (chapter 5) among other places. A major advance was made by Martin, Siggia and Rose in 1973 \cite{martin1973statistical}  (MSR) who realized that one can study classical variables subject to stochastic noise by representing them as Heisenberg operators with appropriately defined commutation relations with {\em conjugate fields}. These observations allowed MSR to write down a generating functional that naturally lends itself to a perturbative treatment and the use of other field theoretic tools. The most important early development that followed was the work by De Dominicis, Peliti and Janssen \cite{dominicis1976techniques,%
janssen1976lagrangean,de1978field}. They realized that the conjugate field that was a fundamental insight and step in the MSR theory, and was introduced in some sense {\em by hand} there, arises naturally if one takes a different point of view. Here one starts the analysis by writing the probability of a path taken by the stochastic variables or fields, which can be used to construct a generating functional expressed as a functional path integral. We refer the reader specifically to Ref.~\cite{jensen1981functional}, which in addition to describing various extensions of the earlier work provides an instructive comparison between the MSR operator based formalism and the functional integral method of De Dominicis, Peliti and Janssen. This latter path integral formalism, which we describe in more detail in this review, has been used to study systems subject to both fast and quenched noise \cite{de1978dynamics}. It paved the way for a more complete understanding of the ensemble averaged dynamics of these systems \cite{sompolinsky1981dynamic,coolen1993dynamics} and, later on, the single sample \cite{biroli1999dynamical,roudi2011dynamical} dynamics of spin glasses and related models. 

Despite the remarkable power of the path integral approach, a pedagogical introduction to the method and its key results is lacking. The aim of this paper is to provide a review to fill this gap. The material presented here has previously been used by the authors in a number of lectures and courses (see e.g.\ the online lectures in \cite{roudi2013}); parts of it can also be found in \cite{fischer1993spin}.

Some of the material has also been reviewed and discussed elsewhere. The diagrammatic approach to the study of Langevin equations has been described in several classical books \cite{ma2000modern,zinn2002quantum,de2006random}, as well as a recent review by Chow and Buice \cite{chow2015path}. The supersymmetric method applied to dynamics is covered in \cite{zinn2002quantum}, for example. In this review we aim to combine these classical results with more novel material, including results on the dynamics of hard spin models, in a coherent and consistent notation. By going into more depth in the calculations, we also hope to provide a useful resource, both for newcomers and more experienced researchers in the field.

In what follows, we start by introducing the path integral formalism for the dynamics of a random variable that evolves according to a stochastic differential equation. We then illustrate how to perform a perturbation analysis and construct Feynman diagrams within this approach. We also show how this treatment can be cast into a supersymmetric form and discuss the physical interpretation of the supersymmetries that are revealed by this approach. Subsequently, we extend the treatment to systems with quenched interactions, especially spin glasses. Finally, we discuss how the path integral formulation can be used to study the dynamics of hard spin models with quenched interactions through a systematic expansion in the strength of the couplings. Before diving into the path integral formulation, however, we review in the next section the two main approaches (Ito and Stratonovich) for interpreting a stochastic differential equation.

\section{Ito vs Stratonovich}

Consider the linear Langevin equation
\be
\dot\phi(t) = -\mu\phi(t) + h(t) + \zeta(t)
\label{lin_Lang}
\ee
where $h(t)$ is a small field used to probe linear response and $\zeta$
is zero mean Brownian or thermal noise with correlation function
$\lav\zeta(t)\zeta(t')\rav = 2T\delta(t-t')$. Eq.~(\ref{lin_Lang}) can of course be
integrated directly from an initial condition at time $t=0$
\be
\phi(t) = \phi(0)e^{-\mu t} + 
\int_0^t\! dt'\, e^{-\mu(t-t')} [\zeta(t')+h(t')]
\ee
and by averaging over the noise and, uncorrelated with it, $\phi(0)$,
we get the correlation function
\be
C(t,t') = \lav\phi(t)\phi(t')\rav = \frac{T}{\mu} e^{-\mu|t-t'|} +
\left( \lav\phi^2(0)\rav - \frac{T}{\mu}\right) e^{-\mu(t+t')}			\label{eq:freecorr}
\ee
and the response function
\be
R(t,t') = \frac{\delta \lav\phi(t)\rav}{\delta h(t')} =
\Theta(t-t') e^{-\mu(t-t')}
\label{response}
\ee
where $\Theta(x)$ is the Heaviside step function, with $\Theta(x)=1$ for $x>0$ and $\Theta(x)=0$ for $x<0$.

We will see below that in the diagrammatic perturbation theory for
treating {\em nonlinear} Langevin equations, we will need the equal
time response $R(t,t)$. This is a priori undefined, however, since the
response function has a step discontinuity at $t'=t$. 

An alternative perspective on this issue is provided by writing the response as a correlation function with the noise. Using the fact
that the average over the distribution of $\zeta$ is with probability
weight $P[\zeta] \sim \exp[-(4T)^{-1}\int\!dt\,\zeta^2(t)]$, and
integrating by parts, we can write
\be
R(t,t') = \lav\frac{\delta \phi(t)}{\delta \zeta(t')}\rav =
\frac{1}{2T} \lav \phi(t)\zeta(t') \rav
\label{response_corr}
\ee
For the equal-time response $R(t,t)$ we therefore require $\lav \phi(t)\zeta(t) \rav$, an equal-time product of fluctuating quantities. 

There are two conventions for assigning values to such quantities, due to Stratonovich and to Ito. We will briefly review these conventions and refer the reader to 
\cite{gardiner1985handbook} (Chapter 4) for a more comprehensive treatment.

\subsection{Stratonovich convention}

The idea of the Stratonovich convention is that, physically, $\zeta$ is a noise process with nonzero correlation
time, so we should really write $\lav \zeta(t)\zeta(t') \rav =
C_\zeta(t-t')$ with $C_\zeta$ an even function, decaying quickly to zero
for $|t-t'|$ greater than some small correlation time $\tau_0$, and
whose integral is $\int\!dt\, C_\zeta(t)=2T$. Then, setting the field
$h$ to zero as it is no longer needed at this point:
\be
\phi(t) = \phi(0)e^{-\mu t} + 
\int_0^t\! dt'\, e^{-\mu(t-t')} \zeta(t')
\ee
so
\be
\lav \phi(t)\zeta(t) \rav = \int_0^t\! dt'\, e^{-\mu(t-t')}
C_\zeta(t-t') \simeq \int_{0}^t \! dt' \, C_\zeta(t-t') \simeq T
\ee
where the second equality follows from the fast decay of $C_\zeta$ compared to the
macroscopic timescales of $\order(1/\mu)$ so that we can approximate
$e^{-\mu(t-t')}=1$. When considering
$\lav \phi(t)\zeta(t') \rav$ with $t>t'$ (more precisely, $t-t'\gg \tau_0$) on
the other hand, all of the ``mass'' of the correlation function is
captured in the integration range; it therefore acts just like a
$\delta$-function and we get $\lav \phi(t)\zeta(t') \rav = 2T
\exp[-\mu(t-t')]$ as expected from\eqq{response}{response_corr}. 

To summarize, in the Stratonovich convention, the equal-time value of the response
function is {\em half} of that obtained in the limit $t'\to t-0$. It
is therefore also called the {\em midpoint rule}; see below.

\subsection{Ito convention}

The Ito convention effectively assumes that the noise $\zeta(t)$ acts
``after $\phi(t)$ has been updated'', so it sets
$\lav\phi(t)\zeta(t)\rav = \lim_{t'\to t+0} \lav\phi(t)\zeta(t')\rav =
0$, and hence also $R(t,t)=0$.

\subsection{Discretization}

We will later look at path integral representations of the dynamics
and so need a discretization of the stochastic process
$\phi(t)$. We will now see that Ito and Stratonovich can be seen as
corresponding to different discretization methods.

Let us discretize time $t=n\Delta$, with $\Delta$ a small time step
eventually to be taken to zero, and write $\phi_n = \phi(t=n\Delta)$
and $h_n = h(t=n\Delta)$.  The noise variables over the interval
$\Delta$ are $\zeta_n = \int_{n\Delta}^{(n+1)\Delta}\!dt\, \zeta(t)$,
with $\lav \zeta_m \zeta_n \rav = 2T\Delta\delta_{mn}$.  Then a suitable
discrete version of\eq{lin_Lang} is
\be
\phi_{n+1}-\phi_n = \Delta[(1-\lambda)(-\mu\phi_n+h_n) +
\lambda(-\mu\phi_{n+1}+h_{n+1})] + \zeta_n
\label{discr_lin}
\ee
for any $\lambda\in[0,1]$; here we are evaluating (the non-noise part
of) the right hand side of\eq{lin_Lang} as a weighted combination of
the values at the two ends of the interval $\Delta$. It is easy to
solve this linear recursion exactly: from
\be
\phi_{n+1}[1+\Delta\lambda\mu] = \phi_n[1+\Delta(\lambda-1)\mu] + 
\Delta[(1-\lambda)h_n + \lambda h_{n+1}] + \zeta_n,
\ee
setting
\be
c = \frac{1+\Delta(\lambda-1)\mu}
{1+\Delta\lambda\mu}
\ee
and assuming the initial condition $\phi_0=0$, we have
\beastar
\phi_n &=& \sum_{m=0}^{n-1} c^{n-m-1} \frac{\Delta[(1-\lambda)h_{m} + \lambda
h_{m+1}] + \zeta_{m}}{1+\Delta\lambda\mu}
\\
\lav\phi_n\rav &=& \frac{\Delta}{1+\Delta\lambda\mu} \left\{\lambda h_n +
\sum_{m=1}^{n-1} \frac{c^{n-m-1}}{1+\Delta\lambda\mu} h_m +
c^{n-1}(1-\lambda)h_0\right\}\ .
\eeastar
From this we can read off the response function
$R_{nm}=\partial\lav\phi_n\rav/\partial (\Delta h_m)$; setting
$n=t/\Delta$, $m=t'/\Delta$ and taking $\Delta\to 0$ we then get for
the continuous time response
\be
R(t,t') = \left\{ \begin{array}{lll}
0 & \mbox{for} & t<t' \\
\lambda & \mbox{for} & t=t' \\
\exp[-\mu(t-t')] & \mbox{for} & t>t'\\
\end{array}
\right.
\ee
The value of $\lambda$ only affects the equal-time response; we see
that $\lambda=1/2$ gives the Stratonovich convention, while
$\lambda=0$ gives Ito. Note that in the discretization\eq{discr_lin},
$\lambda=1/2$ corresponds to evaluating the change in $\phi$ at the
midpoint of the interval between $n$ and $n+1$ (hence ``midpoint
rule''). Ito ($\lambda=0$) on the other hand just evaluates at the
left point. Note also that for different times, $t\neq t'$, the two
discretization schemes are equivalent as expected.

The two conventions for multiplying equal-time fluctuations are, for
the systems we will look at, simply different ways of describing the
{\em same} time evolution $\phi(t)$. Cases where the noise strength is
coupled to $\phi$, such as in $\dot\phi = f(\phi) + g(\phi)\zeta$, are
more serious: a convention for the equal-time product $g(\phi)\zeta$
has to be adopted, and the two conventions here actually give {\em
different} stochastic processes $\phi(t)$; the corresponding
Fokker-Planck equations differ by a nontrivial drift term.

For our simpler cases, Ito vs Stratonovich is basically a matter of
taste. Stratonovich is the more ``physical'' because it corresponds to a
noise process $\zeta$ with small but nonzero correlation time; it also
obeys all the usual rules for transformation of variables etc. Ito is
more obvious from the discretized point of view -- it is very much what one might 
naively program in a simulation. We will also see below that it
can lead to technical simplifications in calculations.

\section{The MSR path integral formulation}
\label{MSR}
Now consider the nonlinear Langevin equation
\be
\dot\phi = f(\phi) + h + \zeta
\label{eq:one_body_model}
\ee
and assume for simplicity that $\phi(0)=0$. It is straightforward to
extend the formalism to systems with several components $\phi_i$; the
inclusion of distributions of initial values is discussed briefly in
Sec.~\ref{sec:example}.  We discretize as in\eq{discr_lin}, abbreviating $f_n=f(\phi_n)$:
\be
\phi_{n+1}-\phi_n = \Delta[(1-\lambda)(f_n+h_n) +
\lambda(f_{n+1}+h_{n+1})] + \zeta_n
\label{discr_nonlin}
\ee
The plan from here is to write down a path integral for this process and
evaluate the effects of nonlinearities in $f(\phi)$
perturbatively. Let $\Delta$ be fixed for now and let $M$ be the
largest value of the index $n$ that we are interested in. Abbreviate $\phi
= (\phi_1\ldots \phi_M)$ and let $\psi
= (\psi_1\ldots \psi_M)$ be a vector of conjugate variables; then the relevant generating
function is
\be
Z[\psi] = \int\! d\phi \ P[\phi] \exp\left(i\sum_{n=1}^M\psi_n\phi_n\right)
\label{Z_def},
\ee
where $P[\phi]$ is the probability of the entire history $\{ \phi_n \}$.  Averages can be computed by differentiation with respect to the $\psi_n$ at $\psi=0$. 

Quenched disorder can also be treated: one
averages the generating function over the quenched disorder before performing the
perturbative expansion. In contrast to the case of equilibrium statistical mechanics, performing the average is unproblematic since, because $P[\phi]$ is a normalized distribution, $Z[0]=1$, independent of the parameters of the model~\cite{de1978dynamics}. We return to this approach in Sec.~\ref{sec:quenched_average}.

The $\phi$ are, from\eq{discr_nonlin}, in one-to-one relation with the
noise variables $\zeta=(\zeta_0\ldots \zeta_{M-1})$. We know the
distribution of the latter:
\be
P(\zeta) = (4\pi T\Delta)^{-M/2} \exp\left[-\frac{1}{4T\Delta}\sum_{n=0}^{M-1}
\zeta_n^2\right]
\ee
and so 
\be
P(\phi) = P(\zeta) J(\phi)
\label{eq:jack}
\ee
where $J(\phi)=|\partial \zeta/\partial \phi|$ is the
Jacobian. Using\eq{discr_nonlin} to express the $\zeta$ in terms of the
$\phi$, 
\be
\zeta_n = \phi_{n+1}-\phi_n - \Delta[(1-\lambda)(f_n+h_n) +
\lambda(f_{n+1}+h_{n+1})]
\label{eta_phi}
\ee
we thus have
\beastar
Z[\psi] &=& (4\pi T\Delta)^{-M/2} \int\! d\phi\ J(\phi)\,\exp\Biggl[
i\sum_{n=1}^M\psi_n\phi_n + \nn
&&{}-{}
\frac{1}{4T\Delta} \sum_{n=0}^{M-1} \left(
\phi_{n+1}-\phi_n - \Delta[(1-\lambda)(f_n+h_n) +
\lambda(f_{n+1}+h_{n+1})]\right)^2 
\Biggr]
\eeastar
The square can be decoupled using conjugate integration variables
$\hph = \hph_0\ldots \hph_{M-1}$:
\beastar
\fl Z[\psi] &=& \int\! \frac{d\phi\,d\hph}{(2\pi)^M}\, J(\phi)\,\exp\Biggl[
i\sum_{n=1}^M\psi_n\phi_n \\
\fl &&{}+\sum_{n=0}^{M-1} 
\left(-T\Delta\hph_n^2 + i\hph_n
\left\{-\phi_{n+1}+\phi_n + \Delta[(1-\lambda)(f_n+h_n) +
\lambda(f_{n+1}+h_{n+1})]\right\}\right)
\Biggr]
\nonumber
\eeastar
Now it is time to evaluate $J(\phi)$. Consider the matrix
$\partial\zeta/\partial\phi$, remembering that the index for $\zeta$
runs from 0 to $M-1$, and the one for $\phi$ from 1 to $M$. The diagonal
elements $\partial \zeta_n/\partial \phi_{n+1}$ are then,
from\eq{eta_phi}, $1 - \Delta\lambda f'_{n+1}$ (with
$f'_n=f'(\phi_n)$); the elements just below the diagonal are $\partial
\zeta_n/\partial\phi_n = -1 - \Delta(1-\lambda) f'_n$. All other
elements are zero; in particular the matrix has all zeros above the
diagonal and so its determinant is the product of the diagonal
elements, giving
\be
J(\phi) = \prod_{n=1}^M (1 - \Delta\lambda f'_n) =
\exp\left[ \sum_{n=1}^M \ln(1 - \Delta\lambda f'_n)\right] =
\exp\left[ -\Delta \lambda \sum_{n=1}^M f'_n\right]
\ee
where the last equality anticipates that $\Delta$ will be made small
so that we can discard $\order(\Delta^2)$ terms in the exponent. Note
that for the Ito convention ($\lambda=0$), $J\equiv 1$ identically, which is one of the reasons for preferring Ito.
The alternative to the above direct evaluation of $J(\phi)$ is to
represent the determinant $|\partial\zeta/\partial\phi |$ as an integral
over Grassmann variables; this will be discussed in Sec.~\ref{StratonovichSUSY}.

With $J(\phi)$ evaluated, we have
\beastar
\fl Z[\psi] &=& \int\! \frac{d\phi\,d\hph}{(2\pi)^M}\, \exp\Biggl[
\sum_{n=1}^M (i\psi_n\phi_n -\Delta \lambda f'_n)
\label{Z_after_noise_average}
\\
\fl &&{}+\sum_{n=0}^{M-1} 
\left(-T\Delta\hph_n^2 + i\hph_n
\left\{-\phi_{n+1}+\phi_n + \Delta[(1-\lambda)(f_n+h_n) +
\lambda(f_{n+1}+h_{n+1})]\right\}\right)
\Biggr]
\nonumber
\eeastar
Defining the average over a (normalized, complex valued) measure 
\be
\langle \ldots \rangle_S = \int\! \frac{d\phi\,d\hph}{(2\pi)^M}\, \ldots \ \exp(-S)
\label{av_def}
\ee
with the ``action''
\be
S = \sum_{n=0}^{M-1} 
\left(T\Delta\hph_n^2 - i\hph_n 
\left\{-\phi_{n+1}+\phi_n + \Delta[(1-\lambda)f_n +
\lambda f_{n+1}]\right\}\right)
+ \Delta \lambda \sum_{n=1}^M f'_n
\label{action}
\ee
one can also write
\be
Z[\psi] = \left \langle \exp\left(i\sum_{n=1}^M \psi_n\phi_n
+i\sum_{n=0}^{M-1} 
\hph_n \Delta[(1-\lambda)h_n + \lambda h_{n+1}]
\right)
\right \rangle_S
\label{Z_av}
\ee
From this representation one has, in particular (taking all
derivatives at $\psi=h=0$, and adopting the convention
$\hph_{-1}=0$) the expressions for correlation and reponse functions
\beastar
C_{nm}&=&\lav\phi_n\phi_m\rav = \frac{\partial}{i\partial \psi_n}
\frac{\partial}{i\partial \psi_m} Z = \langle \phi_n \phi_m\rangle_S
\\
R_{nm}&=& \frac{\partial\lav\phi_n\rav}{\partial (\Delta h_m)} = 
\frac{\partial}{\partial (\Delta h_m)}
\frac{\partial}{i\partial \psi_n} Z = \langle \phi_n
i\{(1-\lambda)\hph_m+\lambda\hph_{m-1}\}\rangle_S
\eeastar
It also follows that averages of any product of $\hph$'s vanish. This is because (as remarked above)
$Z=1$ for $\psi=0$,  whatever value the $h_n$
take. 
As a result, derivatives of any order of $Z$ with respect to $h$ vanish when taken at $\psi=0$. We can combine the expressions for correlation and response if
we define new variables as
\beastar
\ps{n} &=& \phi_n \qquad (n=1\ldots M)
\\
\pss{n} &=& i\{(1-\lambda)\hph_n+\lambda\hph_{n-1}\} \qquad (n=0\ldots
M-1)
\eeastar
Arranging these into a big vector $\eta=(\ps{1}, \ldots, \ps{M},
\pss{0}, \ldots, \p{2\,}{M-1})$ gives
\be
\langle \eta \eta \T \rangle_S = \left(\begin{array}{cc} C & R \\ R\T & 0 \\ \end{array}
\right) = G
\label{G}
\ee
and the calculation of $G$ (the ``propagator'' in quantum field
theory) is often the main goal of the analysis. In the case of nonzero initial conditions one also wants to be able to calculate the means $\langle \eta\rangle$, as discussed in Sec.~\ref{sec:nonzero_fields} below. The fields $\psi$ and $h$ have served
their purpose and will be set to zero from now on.

\section{Perturbation theory}
\label{sec:perturbation}
To illustrate the perturbative expansion, consider a concrete example:
\be
f(\phi) = -\mu \phi - \frac{g}{3!} \phi^3
\label{model}
\ee
(the $3!$ factor is for later convenience). For $g=0$, we recover a
solvable linear Langevin equation; correspondingly, the
action in\eq{action} becomes quadratic. For $g\neq0$, we therefore
separate this quadratic part and write
\beastar
\fl S &=&S_0 + \Si \\
\fl S_0 &=& \sum_{n=0}^{M-1} 
\left(T\Delta\hph_n^2 - i\hph_n 
\left\{-\phi_{n+1}+\phi_n - \Delta\mu[(1-\lambda)\phi_n +
\lambda \phi_{n+1}]\right\}\right)
+\Delta\mu \lambda M
\\
\fl \Si &=& i\frac{g}{6}\Delta\sum_{n=0}^{M-1} 
\hph_n[(1-\lambda)\phi_n^3 + \lambda \phi_{n+1}^3]
-\frac{g}{2}\Delta \lambda \sum_{n=1}^M \phi_n^2
\eeastar
Rearranging the first sum, the non-quadratic (or ``interacting'' in a
field theory) part of the action can be written in the simpler form
\be
\Si = \frac{g}{6}\Delta\sum_{n=0}^{M-1}
\pss{n}\ps{n}^3 - \frac{g}{2} \Delta \lambda \sum_{n=1}^M \ps{n}^2
\label{Sint}
\ee
We have dropped the last term, $\lambda\hph_{M-1}\phi_{M}^3$ from the
first sum; since (by causality) whatever we do with this last term
does not affect any of the results for earlier times, we just have
to make $M$ larger than any times of interest for this omission to be
irrelevant. Similarly, whether we have the sums start at $n=0$ or
$n=1$ has a vanishing effect for $\Delta\to 0$, so in the following we
will leave off the summation ranges.

Now for $g=0$ we have $S=S_0$, and a corresponding normalized {\em
Gaussian} measure over the $\eta$ for which we denote averages by
$\langle \cdots\rangle_0$ and the corresponding propagator by
$G_0=\langle \eta\eta\T\rangle_0$. For $g\neq 0$, our desired averages are then
written as
\be
\langle \cdots \rangle_S = \left \langle \cdots\, \exp(-\Si)\right \rangle_0
\ee
Assuming $g$ and hence $\Si$ to be small, we thus arrive at the
{\em perturbative expansion}

\be
\langle \ldots\rangle_S= \sum_{k=0}^\infty \frac{1}{k!} \left\langle \ldots\, (-\Si)^k \right\rangle_0
\label{perturb}
\ee
which is a series expansion (in general asymptotic, non-convergent) in
the nonlinearity parameter $g$.  We can now evaluate each term in this series using an elementary 
fact about multivariate Gaussian statistics known to physicists as {\em Wick's theorem}: the average 
of any product of Gaussian random variables is found by summing over all possible
pairings; symbolically (leaving out all indices etc)
\be
\langle \eta\eta\ldots \eta\rangle _0 = \sum_{\rm all\ pairings} \langle \eta\eta\rangle _0 \ldots
\langle \eta\eta\rangle _0
\ee
Let us apply this to the simplest example. We know that $\langle 1\rangle_S=1$, so
this should be true to all orders in the expansion\eq{perturb}. Up to $\order(g)$ one has
\bea
\langle 1 \rangle_S &=& \langle 1\rangle_0-\langle\Si\rangle_0 + \ldots
\\
&=& 1 - g\Delta\sum_n\left\langle 
\frac{1}{6} \pss{n}\ps{n}^3 - \frac{1}{2} \lambda \ps{n}^2
\right\rangle_0
\label{one}
\eea
Using Wick's theorem, the fourth-order average is
\be
\langle \pss{n}\ps{n}\ps{n}\ps{n}\rangle_0 = 3\langle \pss{n}\ps{n}\rangle_0 \langle \ps{n}\ps{n}\rangle_0
= 3 R^0_{nn} C^0_{nn}
\ee
(Why the 3? There are 3 different pairings between the 4 four
variables---(1,2) \& (3,4), (1,3) \& (2,4), and (1,4) \& (2,3)---but
all give the same product of averages.) Here one sees how the equal
time response function appears in the formalism. Inserting the value
$R^0_{nn}=\lambda$ that we found earlier, \eq{one} thus becomes
\be
\langle 1 \rangle = \langle 1 \rangle_0 -
 g\Delta\sum_n \left(
\frac{1}{2} \lambda C^0_{nn} - \frac{1}{2} \lambda C^0_{nn} \right)
 = 1 + \order(g^2)
\ee
as it should be. This illustrates how the nontrivial determinant $J(\phi)$
that arises in the Stratonovich formalism ($\lambda=1/2$) and appears
as the term proportional to $\lambda$ in the interaction part of the
action\eq{Sint} is essential for maintaining the correct
normalization. Similar cancellations occur at all orders in $g$. Ito
($\lambda=0$) is simpler here: the terms in the perturbation expansion
of $\langle 1\rangle_S = 1+\ldots$ all vanish {\em individually}, rather than just
cancelling each other out.

\section{Diagrams}
\label{Diags}
\subsection{Basics}

Diagrams are just a pictorial way of keeping track of the various
terms in the perturbation expansion \eref{perturb}, as evaluated using
Wick's theorem. Having illustrated the equivalence between Ito and
Stratonovich above, we stick to Ito ($\lambda=0$) for now, where the
nontrivial part of the action is simply
\be
\Si = \frac{g}{6}\Delta\sum_n\pss{n}\ps{n}^3
\label{Sint_ito}
\ee
To illustrate the diagrammatic notation, consider again the expansion
for the normalization factor
\be
\langle 1 \rangle = \langle 1 \rangle_0-\langle \Si\rangle _0 + \half \langle \Si^2\rangle _0 + \ldots
\ee
We have dealt with the zeroth and first order terms above. The second order term is
\be
\half \left(-\frac{g}{6}\right)^2 \Delta^2 \sum_{m,n} 
\left\langle \pss{m}\ps{m}^3\pss{n}\ps{n}^3\right\rangle _0
\label{ex}
\ee
Represent each of the summed over time indices $m$ and $n$ by a vertex
with four ``legs'' that symbolize the four $\eta$ factors with the
corresponding time index. Each vertex comes with a factor $-g/6$ from $-\Si$. Both of
the time indices are summed over and the result multiplied by
$\Delta$; in the limit $\Delta\to 0$ these scaled sums of course
become time integrals. The time indices are not fixed by our choice
of observable to average, and are therefore called ``internal'' -- we will
see external vertices in a moment. Now, having drawn the vertices, we
can connect the legs in a number of ways; these represent the
different pairings that Wick's theorem gives for the average
in \eref{ex}. E.g.\ the diagram where the legs from both vertices are
connected to legs from the same vertex only
\be
\parbox{35mm}{
\begin{fmfgraph*}(40,25)
\fmfleft{i}
\fmfright{j}
\fmfdot{m,n}
\fmf{phantom}{i,m,n,j}
\fmf{plain}{m,m}
\fmf{plain,left=90}{m,m}
\fmf{plain}{n,n}
\fmf{plain,left=90}{n,n}
\fmflabel{$m$}{m}
\fmflabel{$n$}{n}
\end{fmfgraph*}
}
\ee
represents all the pairings where $\eta_n$'s are connected to
$\eta_n$'s only, and $\eta_m$'s to $\eta_m$'s only. At each vertex
there are three choices for pairings of this form, so this diagram has
the value
\bea
\fl  \half \left(-\frac{g}{6}\right)^2 \Delta^2 \sum_{m,n} 
3\langle \pss{m}\ps{m}\rangle_0 \langle \ps{m}\ps{m}\rangle_0 \times
3\langle \pss{n}\ps{n}\rangle_0 \langle\ps{n}\ps{n}\rangle_0 =\\
\fl \hspace{8cm} \half \left\{\left(-\frac{g}{6}\right)
\Delta \sum_{n} 3\langle \pss{n}\ps{n}\rangle_0 \langle\ps{n}\ps{n}\rangle_0 
\right\}^2\nonumber 
\eea
This illustrates an important fact: if a diagram separates into
subparts which are not connected, its value is just the product of the
two diagrams separately (apart from the overall factor of $1/2$ here,
which comes from the expansion of $\exp(-\Si)$).

The diagram above certainly does not exhaust all the Wick pairings
of\eq{ex}. So what are the other diagrams corresponding to\eq{ex}? We
can have either two $mn$-pairings, one $nn$ and one $mm$; or four $mn$
pairings. Altogether, one would therefore write\eq{ex} in diagrams as:
\be
\mbox{}\!\!\!\!\!\!\!\!\!\!\!\!\!\!\!\!\!\!
\parbox{35mm}{\begin{fmfgraph*}(40,25)
\fmfleft{i}
\fmfright{j}
\fmflabel{$m$}{m}
\fmflabel{$n$}{n}
\fmfdot{m,n}
\fmf{phantom}{i,m,n,j}
\fmf{plain}{m,m}
\fmf{plain,left=90}{m,m}
\fmf{plain}{n,n}
\fmf{plain,left=90}{n,n}
\end{fmfgraph*}}
\ \ +
\ \ \ \ \ 
\parbox{25mm}{\begin{fmfgraph*}(20,20)
\fmfleft{i}
\fmfright{j}
\fmfv{label=$m$,label.angle=-70}{m}
\fmfv{label=$n$,label.angle=-110}{n}
\fmfdot{m,n}
\fmf{phantom}{i,m,n,j}
\fmf{plain,left}{i,m}
\fmf{plain,left}{n,j}
\fmf{plain,right}{i,m}
\fmf{plain,right}{n,j}
\fmf{plain,left=0.5}{m,n}
\fmf{plain,right=0.5}{m,n}
\end{fmfgraph*}}
+
\!\!\!\!
\!\!\!\!
\parbox{20mm}{\begin{fmfgraph*}(40,25)
\fmfleft{i}
\fmfright{j}
\fmfright{j}
\fmflabel{$m$}{m}
\fmflabel{$n$}{n}
\fmfdot{m,n}
\fmf{phantom}{i,m,n,j}
\fmf{plain,left=1,tension=0.15}{m,n}
\fmf{plain,left=0.5,tension=0.15}{m,n}
\fmf{plain,right=1,tension=0.15}{m,n}
\fmf{plain,right=0.5,tension=0.15}{m,n}
\end{fmfgraph*}}
\ee
We write down the value of the last of these: start with
$\pss{m}$. From the diagram, this must be connected to a $\eta$ on the
other vertex, \ie\ either $\pss{n}$ or one of the three $\ps{n}$. In
the first case, each of the remaining $\ps{m}$ legs is connected to a
$\ps{n}$ leg; there are $3\times2\times1=6$ ways of making those
connections (pairings). In the second case, the $\eta_2$ factor from
the second vertex, $\pss{n}$, must be connected to one of the three
$\ps{m}$ vertices, and the remaining two $\ps{m}$ and $\ps{n}$ legs
can be connected to each other in two ways. Thus the value of the
diagram is:
\be
\fl \half (-g/6)^2 \Delta^2 \sum_{m,n} 
\left\{
6\langle \pss{m}\pss{n}\rangle_0 (\langle \ps{m}\ps{n}\rangle_0)^3 + 18
\langle \pss{m}\ps{n}\rangle_0 \langle \ps{m}\pss{n}\rangle_0(\langle \ps{m}\ps{n}\rangle_0)^2
\right\}
\ee
(Exercise: Evaluate the remaining second order diagram and, as a check,
count all pairings. The diagram just above dealt with $6+18=24$
pairings, while the first one corresponded to $3\times 3=9$
pairings. Since we have 8 $\eta$'s in total, there are
$7\times5\times3\times1 = 8!/(4!2!^4) = 105$ pairings overall, hence the
remaining diagram must correspond to to $105-24-9=72$ pairings.)

In terms of diagrams, it is now quite easy to understand that $\langle 1\rangle_S=1$
as it should to all orders in $g$. Consider an arbitrary diagram in
the expansion; let us assume it is connected, otherwise consider the
subdiagrams separately. Now within the Ito convention the response
function $\langle \ps{n}\pss{m}\rangle_0$ is nonzero only for $n>m$ (for
Stratonovich, the value for $n=m$ is also nonzero; for $n<m$ it is
zero either way from causality). So to make the diagram nonzero, we
have to connect the $\pss{n_1}$ from a given vertex, with time index
$n_1$, say, to the $\ps{n_2}$ leg at {\em another} vertex, $n_2$. The
$\pss{n_2}$ from that vertex must in turn be connected to $\ps{n_3}$
on another vertex and so on. Eventually, because we have a finite
number of vertices, we must come back to our original vertex $n_1$. In
the ``ring'' sequence $n_1$, $n_2$, $n_3$, \ldots, $n_1$ there are at
least two time indices that are in the ``wrong'' order for the
response function to be nonzero, so that the diagram contains at least
one vanishing factor and thus vanishes itself.

The moral of the story so far is: diagrams factorize over disconnected sub-diagrams, and 
the sub-diagrams with only internal (summed-over) vertices vanish. The latter are
also called ``vacuum diagrams'' in field theory. 

\subsection{Diagrams for correlator/response; Dyson equation}

Next we look at the diagrams for the propagator, which encapsulates correlation and response functions,
$\langle \p{\alpha}{i}\p{\beta}{j}\rangle_S$ where $\alpha,\beta\in\{1,2\}$. We now
have two ``external'' vertices with fixed time indices $i$ and $j$;
the propagator is represented as a double line between these two
vertices. It is also called the ``full'' or ``dressed'' 
propagator, in contrast to the ``bare'' propagator that is
represented by the single lines in the diagrams and results from the
pairings in Wick's theorem. To zeroth order in $g$, full and bare
propagator are obviously equal. To first order, we have,
from\eq{perturb},
\beastar
\GG{\alpha}{i}{\beta}{j} = \langle\p{\alpha}{i}\p{\beta}{j}\rangle_S
&=&  \langle \p{\alpha}{i}\p{\beta}{j}\rangle_0 -  \langle \p{\alpha}{i}\p{\beta}{j}\Si\rangle_0 +
\ldots \\
&=&  \langle \p{\alpha}{i}\p{\beta}{j}\rangle_0 - (g/6)\Delta\sum_n
 \langle\p{\alpha}{i}\p{\beta}{j}\pss{n}\ps{n}^3\rangle_0 + \ldots
\eeastar
In diagrams, we have two new contributions from the first order term,
depending on whether we pair up $\eta_{i}$ with $\eta_{j}$ or with one
of the $\eta_{n}$:
\be
\parbox{20mm}{\begin{fmfgraph*}(10,20)
\fmfleft{i}
\fmfright{j}
\fmfv{label=$i$,label.angle=-90}{i}
\fmfv{label=$j$,label.angle=-90}{j}
\fmf{dbl_plain}{i,m,n,j}
\end{fmfgraph*}}
=
\ \ \ \ 
\parbox{15mm}{\begin{fmfgraph*}(10,20)
\fmfleft{i}
\fmfright{j}
\fmfv{label=$i$,label.angle=-90}{i}
\fmfv{label=$j$,label.angle=-90}{j}
\fmf{plain}{i,m,n,j}
\end{fmfgraph*}}
+
\ \ \ \ 
\parbox{15mm}{\begin{fmfgraph*}(10,15)
\fmfleft{i}
\fmfright{j}
\fmftop{d1,d2}
\fmfv{label=$i$,label.angle=-90}{i}
\fmfv{label=$j$,label.angle=-90}{j}
\fmflabel{$n$}{n}
\fmfdot{n}
\fmf{plain,left}{d1,n,d2}
\fmf{plain,right}{d1,n,d2}
\fmf{plain}{i,j}
\end{fmfgraph*}}
+
\ \ \ \ \ 
\parbox{15mm}{\begin{fmfgraph*}(10,20)
\fmfleft{i}
\fmfright{j}
\fmfv{label=$i$,label.angle=-90}{i}
\fmfv{label=$j$,label.angle=-90}{j}
\fmfv{label=$n$,label.angle=-90}{n}
\fmfdot{n}
\fmf{plain,tension=0.5}{i,n,n,j}
\end{fmfgraph*}}
+
\ \ldots
\ee
The second of these has a disconnected part with only internal
vertices, so vanishes. The same argument applies to higher
order diagrams as well: we only need to consider {\em connected}
diagrams\footnote{Digression on equilibrium statistical mechanics: equilibrium
averages can be evaluated by a diagrammatic expansion very similar to
the one here. The main difference is that the partition function
(whose perturbative expansion is the same as the one for $\langle 1\rangle_S$ above)
is not automatically normalized. Averages are thus written as
$\langle \ldots\rangle = \langle \ldots e^{-\Si}\rangle_0/\langle e^{-\Si}\rangle_0$ and the denominator,
when expanded, ensures again that disconnected diagrams
vanish. Equivalently, one can think of averages as derivatives of the
{\em log} partitition function; the perturbative
expansion for $\ln Z$ consists of just the {\em connected} diagrams
from the expansion of $Z$.}. Thus, evaluating the surviving diagram,
\be
\GG{\alpha}{i}{\beta}{j} = \GGn{\alpha}{i}{\beta}{j} - (g/6) \Delta\sum_n
\left\{ 3\GGn{\alpha}{i}{1}{n} \GGn{1}{n}{1}{n} \GGn{2}{n}{\beta}{j} +
3\GGn{\alpha}{i}{2}{n} \GGn{1}{n}{1}{n} \GGn{1}{n}{\beta}{j} \right\}
\label{propg}
\ee
If we define a matrix $\Sigma^1$ by
\be
\Sig{\gamma}{m}{\delta}{n}^1 = -(g/2) \Delta^{-1}\delta_{mn} \GGn{1}{n}{1}{n} 
(\delta_{\gamma,1}\delta_{\delta,2}+\delta_{\gamma,2}\delta_{\delta,1})
\label{Sigg}
\ee
and agree to absorb a factor of $\Delta$ into matrix products, so that
\be
(AB)_{\alpha i,\gamma k} = \Delta \sum_{\beta,j} A_{\alpha i,\beta j}
B_{\beta j, \gamma k}
\ee
then we can write\eq{propg} in matrix form simply as
\be
G = G_0 + G_0 \Sigma^1 G_0
\ee
The way factors of $\Delta$ are absorbed here ensures that matrix
multiplications become time integrals in the natural way for
$\Delta\to 0$, while the factor $\Delta^{-1}\delta_{mn}$ in $\Sigma^1$
becomes $\delta(t-t')$.

To first order in $g$, the above is the whole story for the
propagator. Now look at the second order. Among the diagrams we have
ones such as
\be
\parbox{35mm}{\begin{fmfgraph*}(30,20)
\fmfleft{i}
\fmfright{j}
\fmfv{label=$i$,label.angle=-90}{i}
\fmfv{label=$j$,label.angle=-90}{j}
\fmfv{label=$m$,label.angle=-90}{m}
\fmfv{label=$n$,label.angle=-90}{n}
\fmfdot{m,n}
\fmf{plain}{i,m,m,n,n,j}
\end{fmfgraph*}}
\ \ \ \ \ 
\parbox{35mm}{\begin{fmfgraph*}(30,20)
\fmfleft{i}
\fmfright{j}
\fmfv{label=$i$,label.angle=-90}{i}
\fmfv{label=$j$,label.angle=-90}{j}
\fmfv{label=$m$,label.angle=-90}{m}
\fmfv{label=$n$,label.angle=-90}{n}
\fmfdot{m,n}
\fmf{plain}{i,n,n,m,m,j}
\end{fmfgraph*}}
\ee
These two only differ in how the internal vertices are labelled; since
the latter are summed over, we can lump the two diagrams together into
one {\em unlabelled} diagram. This just gives a factor of 2, which
cancels exactly the prefactor $1/2!$ from the second order expansion
of the exponential in\eq{perturb}. Again, the same happens at higher
orders:
at $\order(g^k)$ we have a prefactor of $1/k!$ but also $k$
internal vertices which can be labelled in $k!$ different ways, so the
unlabelled diagram has a prefactor of one. Thus, the unlabelled
diagram
\be
\parbox{35mm}{\begin{fmfgraph*}(30,20)
\fmfleft{i}
\fmfright{j}
%\fmfv{label=$i$,label.angle=-90}{i}
%\fmfv{label=$j$,label.angle=-90}{j}
%\fmfv{label=$m$,label.angle=-90}{m}
%\fmfv{label=$n$,label.angle=-90}{n}
\fmfdot{m,n}
\fmf{plain}{i,m,m,n,n,j}
\end{fmfgraph*}}
\ee
has the value (bearing in mind that the $\eta_2$ components at each of
the internal vertices must be connected either to a $\eta_1$ leg at
the other internal vertex, or to an external vertex, and that there
are three choices at each vertex for which pair of $\eta_1$'s to
connect to each other)
\bea
\fl 3\times 3\times (-g/6)^2 \Delta^2\sum_{mn} \bigl(
& & 
\GGn{\alpha}{i}{1}{m} \GGn{1}{m}{1}{m} \GGn{2}{m}{1}{n}
\GGn{1}{n}{1}{n} \GGn{2}{n}{\beta}{j} + \nn
& & 
\GGn{\alpha}{i}{1}{m} \GGn{1}{m}{1}{m} \GGn{2}{m}{2}{n}
\GGn{1}{n}{1}{n} \GGn{1}{n}{\beta}{j} + \nn
& & 
\GGn{\alpha}{i}{2}{m} \GGn{1}{m}{1}{m} \GGn{1}{m}{1}{n}
\GGn{1}{n}{1}{n} \GGn{2}{n}{\beta}{j} + \nn
& & 
\GGn{\alpha}{i}{2}{m} \GGn{1}{m}{1}{m} \GGn{1}{m}{2}{n}
\GGn{1}{n}{1}{n} \GGn{1}{n}{\beta}{j} \bigr)
\label{dooda}
\eea
Using the definition\eq{Sigg}, one sees that in matrix form this is
simply
\be
G_0 \Sigma^1 G_0 \Sigma^1 G_0
\ee
To make this simple form more obvious we have included in\eq{dooda} the
term in the second line, which vanishes because it contains a zero
$G^0_{22}$ factor. We have an example here of a ``one-particle
reducible'' (1PR) diagram, which can be cut in two by cutting just one
bare propagator line, namely, the one in the middle. The result illustrates
that the value of such diagrams factorizes into the pieces they can be
cut into; e.g.\ the diagram
\be
\fl \parbox{40mm}{\begin{fmfgraph*}(40,20)
\fmfleft{i}
\fmfright{j}
\fmfdot{m,n,o}
\fmf{plain,tension=1}{i,m,m,n,n,o,o,j}
\end{fmfgraph*}}
\ \ \ \ \ 
= 
\ \ \ \ \ 
\parbox{60mm}{\begin{fmfgraph*}(60,20)
\fmfleft{i}
\fmfright{j}
\fmfdot{m,n,o}
\fmf{plain}{i,m1}
\fmf{phantom}{m1,m,m2}
\fmf{phantom}{n1,n,n2}
\fmf{phantom}{o1,o,o2}
\fmf{plain}{m2,n1}
\fmf{plain}{n2,o1}
\fmf{plain}{o2,j}
\fmf{plain,tension=0.6,right=90}{m,m}
\fmf{plain,tension=0.6}{n,n}
\fmf{plain,tension=0.6}{o,o}
\end{fmfgraph*}}
\ee
has the value $G_0 \Sigma^1 G_0 \Sigma^1 G_0 \Sigma^1 G_0$. If we sum
up all the diagrams of this form, we get
\be
\fl G = G_0 + G_0 \Sigma^1 G_0 \Sigma^1 + G_0 \Sigma^1 G_0 \Sigma^1 G_0
+ G_0 \Sigma^1 G_0 \Sigma^1 G_0 \Sigma^1 G_0 + \ldots =
[G_0^{-1} - \Sigma^1]^{-1} 
\ee
or
\be
G^{-1} = G_0^{-1} - \Sigma^1
\label{dyson_prelim}
\ee
The inverses are relative to the appropriate unit element $\mident$
for our redefined matrix multiplication, which has elements $I_{\alpha
i,\beta j} = \Delta^{-1} \delta_{\alpha\beta} \delta_{ij}$. (So
$A^{-1} A = \mident$ means, because of the extra factor $\Delta$ in
the matrix multiplication, that $A^{-1}$ is $\Delta^{-2}$ times the
conventional matrix inverse of $A$.)

The expression\eq{dyson_prelim} is an example of a {\em resummation}:
we have managed to sum up an infinite subseries from among all the
diagrams for the propagator. What if we want to sum up {\em all} the
diagrams?

%:
Again we can classify them into one-particle irreducible (1PI) diagrams, which
cannot be cut in two by cutting a single line, and 1PR diagrams that
factorize into their 1PI components. For example, to second order
\begin{eqnarray}
\fl \parbox{12mm}{\begin{fmfgraph*}(10,20)
\fmfleft{i}
\fmfright{j}
\fmf{dbl_plain}{i,j}
\end{fmfgraph*}}
&=&
\parbox{12mm}{\begin{fmfgraph*}(10,20)
\fmfleft{i}
\fmfright{j}
\fmf{plain}{i,j}
\end{fmfgraph*}}
+\ 
\parbox{12mm}{\begin{fmfgraph*}(10,20)
\fmfleft{i}
\fmfright{j}
\fmfdot{n}
\fmf{plain,tension=0.5}{i,n,n,j}
\end{fmfgraph*}}
+\ 
\parbox{15mm}{\begin{fmfgraph*}(15,20)
\fmfleft{i}
\fmfright{j}
\fmfdot{m,n}
\fmf{plain,tension=0.5}{i,m,m,n,n,j}
\end{fmfgraph*}}
\ +\ 
\parbox{15mm}{\begin{fmfgraph*}(15,20)
\fmfleft{i}
\fmfright{j}
\fmfdot{m,n}
\fmf{plain}{i,m,n,j}
\fmf{plain,left,tension=0}{m,n}
\fmf{plain,right,tension=0}{m,n}
\end{fmfgraph*}}
\ +\ 
\parbox{15mm}{\begin{fmfgraph*}(15,20)
\fmfleft{i}
\fmfright{j}
\fmftop{o}
\fmfdot{m,n}
\fmf{plain}{i,m,j}
\fmffreeze
\fmf{plain,left,tension=0.5}{m,n}
\fmf{plain,right,tension=0.5}{m,n}
\fmf{plain,left,tension=0.5}{o,n}
\fmf{plain,right,tension=0.5}{o,n}
\end{fmfgraph*}}
\ +\ \order(g^3)
\label{eq:1PR2nd}
\\
\fl &=& \left(
\parbox{12mm}{\begin{fmfgraph*}(10,10)
\fmfleft{i}
\fmfright{j}
\fmf{plain}{i,j}
\end{fmfgraph*}}
^{-1}
-\left[
\parbox{12mm}{\begin{fmfgraph*}(10,20)
\fmfleft{i}
\fmfright{j}
\fmfdot{n}
\fmf{plain,tension=0.5}{n,n}
\fmf{phantom,tension=0.5}{i,n}
\fmf{phantom,tension=0.5}{j,n}
\end{fmfgraph*}}
+
\parbox{15mm}{\begin{fmfgraph*}(15,20)
\fmfleft{i}
\fmfright{j}
\fmfdot{m,n}
\fmf{plain}{m,n}
\fmf{phantom}{i,m}
\fmf{phantom}{j,n}
\fmf{plain,left,tension=0}{m,n}
\fmf{plain,right,tension=0}{m,n}
\end{fmfgraph*}}
+
\parbox{15mm}{\begin{fmfgraph*}(15,20)
\fmfleft{i}
\fmfright{j}
\fmftop{o}
\fmfdot{m,n}
\fmf{phantom}{i,m,j}
\fmffreeze
\fmf{plain,left,tension=0.5}{m,n}
\fmf{plain,right,tension=0.5}{m,n}
\fmf{plain,left,tension=0.5}{o,n}
\fmf{plain,right,tension=0.5}{o,n}
\end{fmfgraph*}}
\right]
\right)^{-1} + \ \order(g^3)
\end{eqnarray}
More generally, if we denote the values of the different 1PI diagrams
(defined analogously to $\Sigma^1$, \ie\ without the external $G_0$
legs) by $\Sigma^1$, $\Sigma^2$, \ldots then we can get all possible
(1PI and 1PR) diagrams by ``stringing'' together all possible
combinations of 1PI diagrams:
\be
G = G_0 \sum_{k=0}^\infty \sum_{i_1\ldots i_k} \Sigma^{i_1}G_0 \cdots
\Sigma^{i_k} G_0 = G_0 \sum_{k=0}^\infty (\Sigma G_0)^k =
(G_0^{-1} - \Sigma)^{-1}
\ee
where $\Sigma=\sum_{i=1}^\infty
\Sigma^i$. Hence we see that the full propagator,
with all diagrams summed up, can be written in the general form
\be
G^{-1} = G_0^{-1} - \Sigma
\label{dyson}
\ee
where $\Sigma$, the so-called ``self-energy'', is the sum of all 1PI
diagrams. Eq.\eq{dyson} is called the {\em Dyson equation}. An
alternative form that is often useful is
\be
G_0^{-1} G = \mident + \Sigma G
\label{dysonn}
\ee
Let us write this out in terms of the separate blocks corresponding to
correlation and response functions: We have
\be
G_0 =
\left(\begin{array}{cc} C_0 & R_0 \\ R_0\T & 0 \\ \end{array} \right)
\qquad \hence \qquad
G_0^{-1} =
\left(\begin{array}{cc} 0 & (R_0^{-1})\T 
\\ R_0^{-1} & -R_0^{-1}C_0(R_0^{-1})\T \\ \end{array} \right)
\ee
and $G^{-1}$ has the same structure, so that also the $11$ block of
$\Sigma$ must vanish,
\be
\Sigma = 
\left(\begin{array}{cc} 0 & \Sigma_{12} \\ \Sigma_{12}\T & \Sigma_{22}
\\ \end{array} \right)
\ee
This can also be shown diagrammatically: a nonzero contribution to
$\Sigma_{11}$ would correspond to diagrams where the internal vertices
that make connections to the two external vertices do so via $\eta_1$
legs. This leaves all the $\eta_2$ legs to be connected amongst the
internal vertices, and then the same argument as for vacuum diagrams
can be applied. Writing out\eq{dysonn} we have thus
\bea
(R_0^{-1})\T R\T &=& \mident + \Sigma_{12}R\T \\
0 &=& 0 \\
R_0^{-1} C -   R_0^{-1}C_0 (R_0^{-1})\T R\T &=& \Sigma_{12}\T C +
\Sigma_{22}R\T \\
R_0^{-1} R &=& \mident + \Sigma_{12}\T R
\eea
where we have abused the notation by writing $\mident$ also for the
nonzero $M\times M$ sub-blocks of the original $2M\times2M$ matrix
$\mident$. The first and last of these equations are both equivalent
to
\be
R^{-1} = R_0^{-1} - \Sigma_{12}\T
\label{resp_selfenergy}
\ee
implying that $\Sigma_{12}\T$ acts like a self-energy for the reponse
function.  Rearranging, the components of the Dyson equation reduce to
\bea
R_0^{-1} R &=& \Sigma_{12}\T R + \mident \\
R_0^{-1} C &=& \Sigma_{12}\T C + [R_0^{-1}C_0 (R_0^{-1})\T + \Sigma_{22}]R\T
\label{dyson_comp}
\eea
Using\eq{resp_selfenergy}, the last equation can also be solved
explicitly for $C$ as
\be
C = R[R_0^{-1}C_0 (R_0^{-1})\T + \Sigma_{22}]R\T
\label{dyson_corr}
\ee
In the above we have defined $\Sigma$ to be {\em plus} the sum of
all 1PI diagrams; the opposite sign convention also appears in the literature.

\subsection{Self-consistency}

Of course in general one cannot sum up all the diagrams for the
self-energy.  One must make some approximation. In\eq{dyson_prelim} we
used just the lowest order diagram to approximate $\Sigma$. Can we
easily improve the approximation? Yes, if we replace $G_0$ in the
expression for $\Sigma$ by the full propagator $G$; diagrammatically,
\be
\Sigma \ \ \ =
\parbox{12mm}{\begin{fmfgraph*}(10,20)
\fmfleft{i}
\fmfright{j}
\fmfdot{n}
\fmf{dbl_plain,tension=0.5}{n,n}
\fmf{phantom,tension=0.5}{i,n}
\fmf{phantom,tension=0.5}{j,n}
\end{fmfgraph*}}
\ee
Which diagrams does this correspond to? This is easiest to find out
order by order in $g$. To first order, $\Sigma$ and $G$ are
\be
\Sigma \ \ \ =
\parbox{12mm}{\begin{fmfgraph*}(10,20)
\fmfleft{i}
\fmfright{j}
\fmfdot{n}
\fmf{plain,tension=0.5}{n,n}
\fmf{phantom,tension=0.5}{i,n}
\fmf{phantom,tension=0.5}{j,n}
\end{fmfgraph*}}
\qquad\qquad
\parbox{12mm}{\begin{fmfgraph*}(10,20)
\fmfleft{i}
\fmfright{j}
\fmf{dbl_plain}{i,j}
\end{fmfgraph*}}
 =\ \ 
\parbox{12mm}{\begin{fmfgraph*}(10,20)
\fmfleft{i}
\fmfright{j}
\fmf{plain}{i,j}
\end{fmfgraph*}}
+\ 
\parbox{12mm}{\begin{fmfgraph*}(10,20)
\fmfleft{i}
\fmfright{j}
\fmfdot{n}
\fmf{plain,tension=0.5}{i,n,n,j}
\end{fmfgraph*}}
\ee
Re-inserting $G$ into $\Sigma$ we get its form to second order and
therefore also $G$
\be
\Sigma \ \ \ =
\parbox{12mm}{\begin{fmfgraph*}(10,20)
\fmfleft{i}
\fmfright{j}
\fmfdot{n}
\fmf{plain,tension=0.5}{n,n}
\fmf{phantom,tension=0.5}{i,n}
\fmf{phantom,tension=0.5}{j,n}
\end{fmfgraph*}}
+
\parbox{15mm}{\begin{fmfgraph*}(15,20)
\fmfleft{i}
\fmfright{j}
\fmftop{o}
\fmfdot{m,n}
\fmf{phantom}{i,m,j}
\fmffreeze
\fmf{plain,left,tension=0.5}{m,n}
\fmf{plain,right,tension=0.5}{m,n}
\fmf{plain,left,tension=0.5}{o,n}
\fmf{plain,right,tension=0.5}{o,n}
\end{fmfgraph*}}
\ee
\qquad
\quad
\be
\parbox{12mm}{\begin{fmfgraph*}(10,20)
\fmfleft{i}
\fmfright{j}
\fmf{dbl_plain}{i,j}
\end{fmfgraph*}}
 = \ \ 
\parbox{12mm}{\begin{fmfgraph*}(10,20)
\fmfleft{i}
\fmfright{j}
\fmf{plain}{i,j}
\end{fmfgraph*}}
+\ 
\parbox{12mm}{\begin{fmfgraph*}(10,20)
\fmfleft{i}
\fmfright{j}
\fmfdot{n}
\fmf{plain,tension=0.5}{i,n,n,j}
\end{fmfgraph*}}
+\ 
\parbox{15mm}{\begin{fmfgraph*}(15,20)
\fmfleft{i}
\fmfright{j}
\fmfdot{m,n}
\fmf{plain,tension=0.5}{i,m,m,n,n,j}
\end{fmfgraph*}}
\ +\ 
\parbox{15mm}{\begin{fmfgraph*}(15,20)
\fmfleft{i}
\fmfright{j}
\fmftop{o}
\fmfdot{m,n}
\fmf{plain}{i,m,j}
\fmffreeze
\fmf{plain,left,tension=0.5}{m,n}
\fmf{plain,right,tension=0.5}{m,n}
\fmf{plain,left,tension=0.5}{o,n}
\fmf{plain,right,tension=0.5}{o,n}
\end{fmfgraph*}}
\ee
and then we can iterate to get 
\be
\Sigma \ \ \ =
\parbox{12mm}{\begin{fmfgraph*}(10,20)
\fmfleft{i}
\fmfright{j}
\fmfdot{n}
\fmf{plain,tension=0.5}{n,n}
\fmf{phantom,tension=0.5}{i,n}
\fmf{phantom,tension=0.5}{j,n}
\end{fmfgraph*}}
+
\parbox{15mm}{\begin{fmfgraph*}(15,20)
\fmfleft{i}
\fmfright{j}
\fmftop{o}
\fmfdot{m,n}
\fmf{phantom}{i,m,j}
\fmffreeze
\fmf{plain,left,tension=0.5}{m,n}
\fmf{plain,right,tension=0.5}{m,n}
\fmf{plain,left,tension=0.5}{o,n}
\fmf{plain,right,tension=0.5}{o,n}
\end{fmfgraph*}}
+
\parbox{15mm}{\begin{fmfgraph*}(15,20)
\fmfleft{i}
\fmfright{j}
\fmftop{p1,p2}
\fmfdot{m,n,o}
\fmf{phantom}{i,m,j}
\fmffreeze
\fmf{plain,left=0.5,tension=0.5}{m,n}
\fmf{plain,right=0.5,tension=0.5}{m,o}
\fmf{plain,right=0.5,tension=0.5}{o,n}
\fmf{plain,left,tension=0.5}{n,p1}
\fmf{plain,right,tension=0.5}{n,p1}
\fmf{plain,left,tension=0.5}{o,p2}
\fmf{plain,right,tension=0.5}{o,p2}
\end{fmfgraph*}}
+
\parbox{15mm}{\begin{fmfgraph*}(15,25)
\fmfleft{i}
\fmfright{j}
\fmftop{p}
\fmfdot{m,n,o}
\fmf{phantom}{i,m,j}
\fmffreeze
\fmf{plain,left,tension=0.5}{m,n}
\fmf{plain,right,tension=0.5}{m,n}
\fmf{plain,left,tension=0.5}{o,n}
\fmf{plain,right,tension=0.5}{o,n}
\fmf{plain,left,tension=0.5}{o,p}
\fmf{plain,right,tension=0.5}{o,p}
\end{fmfgraph*}}
+\ \ldots
\ee
So the simple operation of replacing $G_0$ by $G$ effectively sums up
an infinite series of ``tadpole'' diagrams in the expansion for the
self-energy.  This is called ``mean field theory".  It gives exact results
for many models with weak long-ranged interactions; for such models the diagrams that have not
been included are negligible in the thermodynamic limit.   The approximation is also called ``self-consistent one-loop'' or ``Hartree-Fock"'' approximation.

\subsection{Diagrammatic conventions}
\label{sec:arrows}

We have used a particular way of drawing diagrams above that does not distinguish between the physical field $\phi=\eta_1$ and the conjugate or ``response" field $\hat\phi=\eta_2$. This approach has the virtue of making all diagrams look essentially like in a static (equilibrium) $\phi^4$ field theory. It does however mean that each diagram can group a number of terms, involving response or correlation functions depending where on each vertex the $\hat\phi$ variables are located in any given Wick pairing.
If one wants to avoid this one can e.g.\ mark on each vertex the $\hat\phi$-leg by an outgoing arrow. Similarly in the response part of the propagator -- the $\eta_1 \eta_2$ sector in our previous convention -- one would mark the $\hat\phi$ end by an arrow. The arrows then  represent the flow of time because in all contractions of the type $\langle \hat\phi \phi\rangle$ the $\hat\phi$ must be at a time before the $\phi$.

The rules for constructing diagrams are modified only slightly in this more detailed diagrammatic representation: an arrow from a vertex cannot connect back to the same vertex as this would give an equal-time response function, which vanishes in the Ito convention. At each vertex one has to consider the possible choices for how the leg with an arrow can be connected to other vertices or external nodes. Diagrams are non-vanishing only if the arrows give a consistent flow of time, i.e.\ arrows cannot form closed loops nor can two legs with outgoing arrows be connected.

Within the above convention, a propagator {\em without} an arrow on it is always a correlation function. In our example, the diagrammatic expansion up to second order of the correlation function then reads
\begin{eqnarray}
\parbox{12mm}{\begin{fmfgraph*}(10,20)
\fmfleft{i}
\fmfright{j}
\fmf{dbl_plain}{i,j}
\end{fmfgraph*}}
&=& \
\parbox{12mm}{\begin{fmfgraph*}(10,20)
\fmfleft{i}
\fmfright{j}
\fmf{plain}{i,j}
\end{fmfgraph*}}
+\ 
\parbox{12mm}{\begin{fmfgraph*}(10,20)
\fmfleft{i}
\fmfright{j}
\fmfdot{n}
\fmf{plain,tension=0.5}{i,n,n}
\fmf{plain_arrow,tension=0.5}{n,j}
\end{fmfgraph*}}
+\ 
\parbox{12mm}{\begin{fmfgraph*}(10,20)
\fmfleft{i}
\fmfright{j}
\fmfdot{n}
\fmf{plain,tension=0.5}{j,n,n}
\fmf{plain_arrow,tension=0.5}{n,i}
\end{fmfgraph*}}
+\ 
\parbox{15mm}{\begin{fmfgraph*}(15,20)
\fmfleft{i}
\fmfright{j}
\fmfdot{m,n}
\fmf{plain,tension=0.5}{i,m,m}
\fmf{plain,tension=0.5}{n,n}
\fmf{plain_arrow,tension=0.5}{m,n,j}
\end{fmfgraph*}}
+\ 
\parbox{15mm}{\begin{fmfgraph*}(15,20)
\fmfleft{i}
\fmfright{j}
\fmfdot{m,n}
\fmf{plain,tension=0.5}{j,n,n}
\fmf{plain,tension=0.5}{m,m}
\fmf{plain_arrow,tension=0.5}{n,m,i}
\end{fmfgraph*}}
+\ 
\parbox{15mm}{\begin{fmfgraph*}(15,20)
\fmfleft{i}
\fmfright{j}
\fmfdot{m,n}
\fmf{plain,tension=0.5}{m,m,n,n}
\fmf{plain_arrow,tension=0.5}{m,i}
\fmf{plain_arrow,tension=0.5}{n,j}
\end{fmfgraph*}}
\nonumber\\
&&{}+\ 
\parbox{15mm}{\begin{fmfgraph*}(15,20)
\fmfleft{i}
\fmfright{j}
\fmfdot{m,n}
\fmf{plain_arrow}{m,n}
\fmf{plain,left,tension=0}{m,n}
\fmf{plain,right,tension=0}{m,n}
\fmf{plain}{m,i}
\fmf{plain_arrow}{n,j}
\end{fmfgraph*}}
\ +\ 
\parbox{15mm}{\begin{fmfgraph*}(15,20)
\fmfleft{i}
\fmfright{j}
\fmfdot{m,n}
\fmf{plain_arrow}{n,m}
\fmf{plain,left,tension=0}{m,n}
\fmf{plain,right,tension=0}{m,n}
\fmf{plain_arrow}{m,i}
\fmf{plain}{n,j}
\end{fmfgraph*}}
\ +\ 
\parbox{15mm}{\begin{fmfgraph*}(15,20)
\fmfleft{i}
\fmfright{j}
\fmfdot{m,n}
\fmf{plain}{m,n}
\fmf{plain,left,tension=0}{m,n}
\fmf{plain,right,tension=0}{m,n}
\fmf{plain_arrow}{m,i}
\fmf{plain_arrow}{n,j}
\end{fmfgraph*}}
\nonumber\\
&&{}+\ 
\parbox{15mm}{\begin{fmfgraph*}(15,25)
\fmfleft{i}
\fmfright{j}
\fmftop{o}
\fmfdot{m,n}
\fmf{plain}{i,m}
\fmf{plain_arrow}{m,j}
\fmffreeze
\fmf{plain,left,tension=0.25}{m,n}
\fmf{plain_arrow,left,tension=0.25}{n,m}
%\fmf{plain_arrow}{n,m}
\fmf{plain,left,tension=0.5}{o,n}
\fmf{plain,right,tension=0.5}{o,n}
\end{fmfgraph*}}
\iffalse
{\begin{fmfgraph*}(15,20)
\fmfleft{i}
\fmfright{j}
\fmftop{o}
\fmfdot{m,n}
\fmf{plain}{i,m}
\fmf{plain_arrow}{m,j}
\fmffreeze
\fmf{plain,left,tension=0.05}{m,n}
%\fmf{plain,right,tension=0.5}{m,n}
\fmf{plain_arrow}{n,m}
\fmf{plain,left,tension=0.5}{o,n}
\fmf{plain,right,tension=0.5}{o,n}
\end{fmfgraph*}}
\fi
\ +\ 
\parbox{15mm}{\begin{fmfgraph*}(15,25)
\fmfleft{i}
\fmfright{j}
\fmftop{o}
\fmfdot{m,n}
\fmf{plain_arrow}{m,i}
\fmf{plain}{j,m}
\fmffreeze
\fmf{plain,left,tension=0.25}{m,n}
\fmf{plain_arrow,left,tension=0.25}{n,m}
%\fmf{plain_arrow}{n,m}
\fmf{plain,left,tension=0.5}{o,n}
\fmf{plain,right,tension=0.5}{o,n}
\end{fmfgraph*}}
\iffalse
{\begin{fmfgraph*}(15,20)
\fmfleft{i}
\fmfright{j}
\fmftop{o}
\fmfdot{m,n}
\fmf{plain}{m,j}
\fmf{plain_arrow}{m,i}
\fmffreeze
\fmf{plain,left,tension=0.05}{m,n}
%\fmf{plain,right,tension=0.5}{m,n}
\fmf{plain_arrow}{n,m}
\fmf{plain,left,tension=0.5}{o,n}
\fmf{plain,right,tension=0.5}{o,n}
\end{fmfgraph*}}
\fi
\ +\ \order(g^3)
\label{eq:corrdiag}
\end{eqnarray}
while for the response function there are fewer diagrams that contribute:
\begin{eqnarray}
\parbox{12mm}{\begin{fmfgraph*}(10,20)
\fmfleft{i}
\fmfright{j}
\fmf{dbl_plain_arrow}{i,j}
\end{fmfgraph*}}
&=& \
\parbox{12mm}{\begin{fmfgraph*}(10,20)
\fmfleft{i}
\fmfright{j}
\fmf{plain_arrow}{i,j}
\end{fmfgraph*}}
+\ 
\parbox{12mm}{\begin{fmfgraph*}(10,20)
\fmfleft{i}
\fmfright{j}
\fmfdot{n}
\fmf{plain,tension=0.5}{n,n}
\fmf{plain_arrow,tension=0.5}{i,n,j}
\end{fmfgraph*}}
+\ 
\parbox{15mm}{\begin{fmfgraph*}(15,20)
\fmfleft{i}
\fmfright{j}
\fmfdot{m,n}
\fmf{plain,tension=0.5}{m,m}
\fmf{plain,tension=0.5}{n,n}
\fmf{plain_arrow,tension=0.5}{i,m,n,j}
\end{fmfgraph*}}
+\ 
\parbox{15mm}{\begin{fmfgraph*}(15,25)
\fmfleft{i}
\fmfright{j}
\fmftop{o}
\fmfdot{m,n}
\fmf{plain_arrow}{i,m}
\fmf{plain_arrow}{m,j}
\fmffreeze
\fmf{plain,left,tension=0.25}{m,n}
\fmf{plain_arrow,left,tension=0.25}{n,m}
%\fmf{plain_arrow}{n,m}
\fmf{plain,left,tension=0.5}{o,n}
\fmf{plain,right,tension=0.5}{o,n}
\end{fmfgraph*}}
\iffalse
{\begin{fmfgraph*}(15,20)
\fmfleft{i}
\fmfright{j}
\fmftop{o}
\fmfdot{m,n}
\fmf{plain_arrow}{i,m}
\fmf{plain_arrow}{m,j}
\fmffreeze
\fmf{plain,left,tension=0.05}{m,n}
%\fmf{plain_arrow,left,tension=0.25}{n,m}
\fmf{plain_arrow}{n,m}
\fmf{plain,left,tension=0.5}{o,n}
\fmf{plain,right,tension=0.5}{o,n}
\end{fmfgraph*}}
\fi
\ +\ \order(g^3)
\label{eq:respdiag}
\end{eqnarray}

Note that the expansions \eref{eq:corrdiag} and \eref{eq:respdiag} are written as the direct analogues of \eref{eq:1PR2nd} and differ from the latter only through the addition of the appropriate arrows. More compact expressions can be obtained in terms of the self-energy. For the correlation function specifically, the identity \eref{dyson_corr} may be more efficient for use in practical calculations. This is illustrated in the example in Sec. \ref{sec:example} below, see \eref{csol} there.
%\peter{two-bubble-diagrams in two equations have been updated so replace the two entire equations by these new versions}
\iffalse
Note: the two-bubble diagrams are a bit awkward because of the arrow in the loop between the internal vertices; an alternative would be to draw these like
\[
 \parbox{15mm}{\begin{fmfgraph*}(15,25)
\fmfleft{i}
\fmfright{j}
\fmftop{o}
\fmfdot{m,n}
\fmf{plain_arrow}{i,m}
\fmf{plain_arrow}{m,j}
\fmffreeze
\fmf{plain,left,tension=0.25}{m,n}
\fmf{plain_arrow,left,tension=0.25}{n,m}
%\fmf{plain_arrow}{n,m}
\fmf{plain,left,tension=0.5}{o,n}
\fmf{plain,right,tension=0.5}{o,n}
\end{fmfgraph*}}
\]
but I'm not sure that's better. Views welcome. I'm aware I need to sort out / delete the vertex labelling above.
\fi

\subsection{Nonzero fields and initial values}
\label{sec:nonzero_fields}

So far we have discussed dynamics without applied fields $h_n$ and with zero initial value $\phi_0$. These restrictions can be lifted, as we now explain. We continue to use the Ito convention ($\lambda=0$). It suffices to discuss nonzero fields, because a nonzero initial value $\phi_0 = c$ can be produced by setting $h_0=c/\Delta$. This gives $\phi_1=c + \order(\Delta^{1/2})$, which in the continuous time limit $\Delta\to 0$ approaches $c$. In the same limit the difference between $\phi_0$ and $\phi_1$ is immaterial, so the above choice of $h_0$ effectively fixes a nonzero initial condition.

For the perturbative approach to work, we need a quadratic action $S_0$ as our baseline while all non-quadratic terms are contained in the interacting part $\Si$. If the field $h_n$ is nonzero, this generates linear terms in the action. We then have two choices: either we include these linear terms in $\Si$ and treat them perturbatively, or we transform variables by expanding the full action
around a stationary or saddle point.

The first approach is relatively straightforward: one now has an extra vertex, with only one $\hat\phi$-leg, and the perturbative expansion is jointly in $g$ and the field amplitude. For the propagator all contributing diagrams must have a total number of legs on the internal vertices that is even so the number of field vertices must also be even, hence the expansion is effectively in $g$ and $h^2$. At $\order(h^2)$ one has the extra diagram

\begin{equation}
\parbox{12mm}{\begin{fmfgraph*}(10,20)
\fmfleft{i}
\fmfright{j}
\fmfdot{m,n}
\fmf{plain}{i,m}
\fmf{plain}{n,j}
\fmf{phantom}{m,n}
\end{fmfgraph*}}
\label{order_h2}
\end{equation}
where each dot with a {\em single} connection is an $h$-vertex,\footnote{It is common in the literature to use a different symbol, such as a cross, to represent the external field.  Here, in the interest of a more uniform notation, we stick with dots, and the number of lines connected at a dot indicates whether it means $h$, $g$, or a coupling of some still-different order, such as the cubic vertex in (\ref{df}).} while at $\order(h^2 g)$ one gets
\begin{eqnarray}
\parbox{17mm}{\begin{fmfgraph*}(15,20)
\fmfleft{i}
\fmfright{j}
\fmfdot{l,m,n}
\fmf{plain,tension=0.5}{i,l,l,m}
\fmf{plain}{n,j}
\fmf{phantom}{m,n}
\end{fmfgraph*}}
+ \ 
\parbox{17mm}{\begin{fmfgraph*}(15,20)
\fmfleft{i}
\fmfright{j}
\fmfdot{l,m,n}
\fmf{plain,tension=0.5}{j,l,l,m}
\fmf{plain}{n,i}
\fmf{phantom}{m,n}
\end{fmfgraph*}}
+ \ 
\parbox{12mm}{\begin{fmfgraph*}(10,10)
\fmfleft{i1,i,i2}
\fmfright{j1,j,j2}
\fmf{plain}{i,m,j}
\fmf{phantom}{i2,i3}
\fmf{plain}{i3,m}
\fmf{phantom}{j2,j3}
\fmf{plain}{j3,m}
%\fmf{phantom,tension=0.5}{m,l,n,m}
\fmf{phantom}{i1,i0,m}
\fmf{phantom}{j1,j0,m}
\fmfdot{i3,j3,m}
\end{fmfgraph*}}
\label{order_h2_g}
\end{eqnarray}
%
\iffalse
\begin{eqnarray}
\parbox{12mm}{\begin{fmfgraph*}(10,20)
\fmfleft{i}
\fmfright{j}
\fmf{dbl_plain}{i,j}
\end{fmfgraph*}}
&=& \ 
\parbox{12mm}{\begin{fmfgraph*}(10,20)
\fmfleft{i}
\fmfright{j}
\fmf{plain}{i,j}
\end{fmfgraph*}}
+\ 
\parbox{12mm}{\begin{fmfgraph*}(10,20)
\fmfleft{i}
\fmfright{j}
\fmfdot{m,n}
\fmf{plain}{i,m}
\fmf{plain}{n,j}
\fmf{phantom}{m,n}
\end{fmfgraph*}}
+\ 
\parbox{12mm}{\begin{fmfgraph*}(10,20)
\fmfleft{i}
\fmfright{j}
\fmfdot{n}
\fmf{plain,tension=0.5}{i,n,n,j}
\end{fmfgraph*}}
\end{eqnarray}
\fi
We could represent the fact that all field vertices have a single $\hat\phi$-leg by an arrow pointing away from the vertex (see Sec.~\ref{sec:arrows}). For the response function part of the propagator there would be an arrow pointing away from one of the external vertices as well so the $\order(h^2)$ diagram in \eref{order_h2} vanishes as it contains an edge with opposing arrows. To a correlator $C(t,t')$ the diagram contributes in the continuous time limit
\begin{equation}
\int dt_1 dt_2\, R_0(t,t_1)h(t_1)\,R_0(t',t_2)h(t_2)
\end{equation}
The higher order diagrams can be evaluated similarly.

In the presence of a field even the mean $\langle \phi_n \rangle$ is in general nonzero. The perturbative expansion for this consists of diagrams with one external vertex and therefore requires an odd number of internal field vertices. The contributions to $\order(h)$ and $\order(hg)$ are
\begin{eqnarray}
\parbox{12mm}{\begin{fmfgraph*}(10,20)
\fmfleft{i}
\fmfright{j}
\fmfdot{n}
\fmf{dbl_plain}{i,n}
\fmf{phantom}{n,j}
\end{fmfgraph*}}
&=& \ 
\parbox{12mm}{\begin{fmfgraph*}(10,20)
\fmfleft{i}
\fmfright{j}
\fmfdot{n}
\fmf{plain}{i,n}
\fmf{phantom}{n,j}
\end{fmfgraph*}}
+\ 
\parbox{12mm}{\begin{fmfgraph*}(10,20)
\fmfleft{i}
\fmfright{j}
\fmfdot{l,j}
\fmf{plain,tension=0.5}{i,l,l,j}
%\fmf{plain}{n,j}
%\fmf{phantom}{m,j}
\end{fmfgraph*}}
%\
%+ \ldots
\end{eqnarray}
One sees that most of the new diagrams \eref{order_h2} and \eref{order_h2_g} in the propagator are products of diagrams like these. In fact one can convince oneself that if one considers the {\em connected} propagator, defined as 
$\langle \eta_{\alpha i}\eta_{\beta j} \rangle - 
\langle \eta_{\alpha i}\rangle \langle \eta_{\beta j} \rangle$, then as in the case without a field only the connected diagrams remain \cite{lancaster2014quantum}; the lowest order one of these is the third diagram in (\ref{order_h2_g}).

Next we look at the saddle point approach, where one defines new variables relative to a stationary point of the action, thus eliminating any linear terms in the transformed action. Looking at (\ref{Z_after_noise_average}) with $\psi_n=0$, the saddle point conditions with respect to $\phi_n$ and $\hat\phi_n$ give
\begin{eqnarray*}
0&=&\hat\phi_n-\hat\phi_{n-1}+\hat\phi_n\Delta f_n' \\
0 &=& -2T \hat\phi_n + i[-\phi_{n+1}+\phi_n+\Delta(f_n+h_n)]
\end{eqnarray*}
The ``initial'' (at $n=M-1$) condition for the first of these equations, which results from stationarity with respect to $\phi_M$, is $\hat\phi_{M-1}=0$, and solving backwards in time then shows that $\hat\phi_n=0$ for all $n$ at the saddle point. With this the second equation is just
\begin{equation}
\phi_{n+1}-\phi_n=\Delta(f_n+h_n)
\end{equation}
which is the discrete time version of the deterministic (noise-free) time evolution
$\dot\phi = f(\phi) + h$.
\newcommand{\phist}{\phi^*}
\newcommand{\fst}{f^*}
\newcommand{\dphi}{\delta\phi}
\newcommand{\df}{\delta f}
If we call the solution of this $\phist$, then we need to use as new variables $\dphi=\phi-\phist$. In terms of these, by making use of the saddle point conditions, the (Ito) action reads
\be
S = \sum_{n=0}^{M-1} 
\left\{T\Delta\hph_n^2 - i\hph_n 
\left[-\dphi_{n+1}+\dphi_n + \Delta \df_n\right]\right\}
\label{action_nonzero_h}
\ee
The difference $\df_n = f_n-\fst_n=f(\phist_n+\dphi_n)-f(\phist_n)$ needs to be expanded in $\dphi_n$ to read off the resulting vertices in the interacting part of the action. For our example (\ref{model}), 
\begin{equation}
\df = -\mu \dphi - \frac{g}{3!}(
3\phist{}^2\dphi
+ 3\phist\dphi^2 
+\dphi^3)
\label{df}
\end{equation}
If we include the $-\mu\dphi$ term in the unperturbed action $S_0$ then the latter has the same form as originally except for the change of variable from $\phi$ to $\dphi$, and the same diagrammatic expansion technique can be applied. The difference is that the interaction part $\Si$ now contains three different kinds of vertices resulting from the terms proportional to $g$ in (\ref{df}), with two, three and four legs respectively and different time-dependent prefactors from the time dependence of $\phist$.

One point to bear in mind is that even though we have defined $\dphi$ relative to the saddle point solution, its average is not in general zero. To $\order(g)$, for example, one has for $\langle \dphi(t)\rangle$ the diagram
\begin{eqnarray}
\parbox{12mm}{\begin{fmfgraph*}(10,20)
\fmfleft{i}
\fmfright{j}
\fmfdot{n}
\fmf{dbl_plain}{i,n}
\fmf{phantom}{n,j}
\end{fmfgraph*}}
&=& 
\ \ \ 
\parbox{12mm}{\begin{fmfgraph*}(10,20)
\fmfleft{i}
\fmfright{j}
\fmfdot{l}
\fmf{plain}{i,l}
\fmf{plain,tension=0.5,left}{l,j}
\fmf{plain,tension=0.5,right}{l,j}
%\fmf{plain}{n,j}
%\fmf{phantom}{m,j}
\end{fmfgraph*}}
%\
%+ \ldots
\end{eqnarray}
coming from the $-(g/2)\phist\dphi^2$ term in\eq{df}, which evaluates to
\be
\langle \dphi(t) \rangle = -\frac{g}{2}\int dt'\, R_0(t,t')\phist(t') C_0(t',t')
\ee

\section{A one-particle example}
\label{sec:example}

Let us apply the diagrammatic formalism developed above to our one-particle model\eq{model}. We want to take a
distribution over the initial values $\phi(0)$ into account here. But this
only takes a small extension of the formalism: if the initial
distribution is a zero mean Gaussian, we can simply include the
average over $\phi(0)$ in the unperturbed average $\langle\ldots\rangle_0$; the
measure is still Gaussian, so we can apply all of the above formalism
except that the values of $C_0$ and $R_0$, i.e.\ the components of the bare
propagator are affected by the presence of uncertainty in the initial
condition. If the distribution has non-Gaussian parts, we include the
Gaussian part as above and the remainder is put into the nontrivial
part of the action $\Si$, giving a new kind of vertex in the
diagrams. A nonzero mean in a Gaussian initial distribution would also be put into $\Si$.

Now let us write down the Dyson equation for our model. Having derived
all relations previously so that they have the obvious limits for
$\Delta\to 0$, we work directly with continuous times. The bare
response function is, from\eq{response},
\be
R_0(t,t') = \Theta(t-t') e^{-\mu(t-t')}
\ee
The inverse of $R_0$ is the operator $\partial_t + \mu$,
since when applied to $R_0$ it gives $\delta(t-t')$.  The other
quantity we need to write down the Dyson equation\eq{dyson_comp} is
$R_0^{-1}C_0 (R_0^{-1})\T$. To find this, it is easiest to start from
the fact that
\be
\phi(t) = \int_{0}^\infty \!\! dt_1\, R_0(t,t_1)[\zeta(t_1) +
\phi(0)\delta(t_1)]
\ee
The lower boundary of the integral here is meant as $0-\epsilon$; the
same applies to all integrals that follow. Averaging the product
$\phi(t)\phi(t')$ gives
\be
C_0(t,t') = \int_{0}^\infty \!\! dt_1\, dt_2\, 
R_0(t,t_1)[2T\delta(t_1-t_2) + 
\lav\phi^2(0)\rav\delta(t_1)\delta(t_2)]R_0(t',t_2)
\ee
so $C_0 = R_0 M R_0\T$ with $M(t_1,t_2)$ given by the square brackets;
thus
\be
(R_0^{-1}C_0 (R_0^{-1})\T)(t,t') = 
2T\delta(t-t') + \lav\phi^2(0)\rav\delta(t)\delta(t')
\ee
Now we can write down the Dyson equation\eq{dyson_comp}:
\bea
\fl \left(\frac{\partial}{\partial t} + \mu\right)
R(t,t') &=& \int_0^\infty \!\! dt''\, \Sigma_{12}(t'',t)R(t'',t') +
\delta(t-t')
\label{dyson_expl}
\\
\fl \left(\frac{\partial}{\partial t} + \mu\right)
C(t,t') &=& \int_0^\infty \!\! dt''\, \Sigma_{12}(t'',t)C(t'',t')
\\ 
\fl & & {} + {} \int_0^\infty \!\! dt''\, [2T\delta(t-t'') +
\lav\phi^2(0)\rav\delta(t)\delta(t'') + \Sigma_{22}(t,t'')]R(t',t'')
\nonumber
\eea
To first order in $g$ the self-energy is, from\eq{Sigg},
\be
\Sigma_{12}(t,t') = -(g/2) \delta(t-t') C_0(t,t) \qquad
\Sigma_{22}(t,t') = 0
\label{Sigt}
\ee
Within this approximation, the Dyson equation becomes
\bea
\fl \left(\frac{\partial}{\partial t} + \mu\right)
R(t,t') &=& -(g/2)C_0(t,t) R(t,t') + \delta(t-t')
\\
\fl \left(\frac{\partial}{\partial t} + \mu\right)
C(t,t') &=& -(g/2)C_0(t,t) C(t,t')
+ 2T R(t',t) + \lav\phi^2(0)\rav\delta(t)R(t',0)
\eea
From the first equation, 
\be
R(t,t') = \Theta(t-t')\exp\left[\mu(t-t')-(g/2)\int_{t'}^t \!\!dt''\,
C_0(t'',t'')\right]
\ee
and with this the second equation for $C$ can also be solved
(compare\eq{dyson_corr}) to give
\be
C(t,t') = \lav\phi^2(0)\rav R(t,0)R(t',0)
+ 2T \int_{t'}^t \!\!dt''\, R(t,t'')R(t',t'')
\label{csol}
\ee
For the simplest case where $C_0$ is time-translation invariant,
corresponding to $\lav\phi^2(0)\rav = C_0(t,t) = T/\mu$, we see that
the effect of $g$ in the response function $R$ is just to replace $\mu
\to \mu + gT/(2\mu)$. At long times the effect on $C$ is similar
though at short times $C$ will not be time-translation invariant.

If me make our first order (one-loop) approximation self-consistent,
the only change is to replace $C_0$ by $C$ in\eq{Sigt} and
correspondingly in the Dyson equation, so that
\be
R(t,t') = \Theta(t-t')\exp\left[-\mu(t-t')-(g/2)\int_{t'}^t \!\!dt''\,
C(t'',t'')\right]
\label{rsol}
\ee
This has a simple interpretation: it corresponds to replacing our
original nonlinear force term\eq{model} by
\be
f(\phi) \approx -\mu \phi - \frac{g}{3!} 3\lav\phi^2\rav \phi
\ee
with $\lav\phi^2(t)\rav = C(t,t)$ to be determined
self-consistently. [The non-self-consistent version instead sets
$\lav\phi^2(t)\rav=C_0(t,t)$.] Assuming that we can find a solution
with $C(t,t)=c=$ constant, we get for $C$ by inserting\eq{rsol}
into\eq{csol} and setting $\tilde\mu = \mu + c g/2$
\be
C(t,t') = \lav\phi^2(0)\rav e^{-\tilde\mu(t+t')} 
+ \frac{T}{\tilde\mu} [e^{-\tilde\mu |t-t'|} - e^{-\tilde\mu (t+t')}]
\ee
For $\lav\phi^2(0)\rav=T/\tilde\mu$ we then indeed get a time-translationally
invariant $C(t,t')= (T/\tilde\mu) \exp(-\tilde\mu |t-t'|)$. This has
$C(t,t)=T/\tilde\mu$ and so the self-consistent equation determining
$c$ and $\tilde\mu$ is
\be
\frac{T}{\tilde\mu} = c \qquad \hence \qquad
\tilde\mu = \mu + \frac{gT}{2\tilde\mu}
\ee
This is the self-consistent analogue of our previous result $\tilde\mu=\mu+(gT)/(2\mu)$, to which it reduces when expanded to first order in $g$.

We see that for our particular model
the self-consistent approximation gives a more sensible result than
the ``vanilla'' first order approximation: it allows a
time-translation invariant solution for both $C$ and $R$
\bea
R(t,t') &=& \Theta(t-t') e^{-\tilde\mu(t-t')} \\
C(t,t') &=& \frac{T}{\tilde\mu} \, e^{-\tilde\mu |t-t'|}
\eea
which also obeys the fluctuation-dissipation theorem (FDT), $R(t,t') = (1/T)(\partial/\partial t') C(t,t')$
for $t>t'$.

\subsection{Mode-coupling theory}
\label{subsec:MCT}

What if we want to improve the approximation to the self-energy
further? The systematic approach is to include the lowest-order
diagram not so far taken into account. We have the only first-order
diagram already; the second-order ``tadpole'' diagrams are also taken
into account through self-consistency. The only missing second order
diagram is therefore the ``watermelon'' diagram
\be
\parbox{25mm}{\begin{fmfgraph*}(25,20)
\fmfleft{i}
\fmfright{j}
\fmfdot{m,n}
\fmfv{label=$m$,label.angle=-135}{m}
\fmfv{label=$n$,label.angle=-45}{n}
%\fmf{plain}{i,m,n,j}
\fmf{phantom}{i,m}
\fmf{phantom}{n,j}
\fmf{plain}{m,n}
\fmf{plain,left,tension=0}{m,n}
\fmf{plain,right,tension=0}{m,n}
\end{fmfgraph*}}
\label{eq:watermelon}
\ee
To work out what contribution to $\Sigma$ this gives, let us revert
temporarily to discrete time notation and label the left and right
vertex $m$ and $n$, respectively. The elements $\Sig{1}{m}{2}{n}$ of
$\Sigma_{12}$ correspond to those pairings where a $\ps{m}$ leg from
vertex $m$ is attached to an external vertex that would be on the left, and the
$\pss{n}$ leg from vertex $n$ attached to an external vertex on the
right. Internally (among the remaining legs) we thus have two
$\ps{m}\ps{n}$ pairings and one $\pss{m}\ps{n}$ pairing. To work out
the prefactor of the diagram, note that there are three choices for
the externally attached $\ps{m}$; there are three choices for which of
the $\ps{n}$ to pair up with $\pss{m}$; and two more choices for how
to make the two remaining $\ps{m}\ps{n}$ pairings. Thus, the diagram
gives for $\Sigma_{12}$
\be
\Sig{1}{m}{2}{n} = 3\times 3\times 2(-g/6)^2 (C^0_{mn})^2 R^0_{nm} =
(g^2/2) (C^0_{mn})^2 R^0_{nm}
\ee
For $\Sigma_{22}$, we have both the $\pss{m}$ and $\pss{n}$ legs
attached externally, and 6 choices for how to connect the three
$\ps{m}$ and $\ps{n}$ legs internally, giving
\be
\Sig{2}{m}{2}{n} = 6(-g/6)^2 (C^0_{mn})^3 = (g^2/6) (C^0_{mn})^3
\ee
We can again sum an infinite series of additional diagrams by
replacing bare ($C_0, R_0$) by full quantities ($C, R$)
here. Reverting to continuous time notation and including the
first-order contribution in $\Sigma_{12}$, we thus get for the
self-energy in the self-consistent two-loop approximation
\bea
\Sigma_{12}(t,t') &=&
 -(g/2) \delta(t-t') C(t,t) + (g^2/2) C^2(t,t') R(t',t) \label{eq:MCTsigma12} \\
\Sigma_{22}(t,t') &=& (g^2/6) C^3(t,t')   \label{eq:MCTsigma22}
\eea

A final comment: up to the order which we have considered, the free
energy components $\Sigma_{12}(t,t')$ and $\Sigma_{22}(t,t')$ are
simple {\em functions} of the correlation and response functions. This
is not normally true once higher order diagrams are taken into
account. For example, if we went to third order in $g$ we would have
to include the diagram
\be
\parbox{25mm}{\begin{fmfgraph*}(25,20)
\fmfleft{i}
\fmfright{j}
\fmfdot{m,n,o}
\fmf{phantom}{i,m}
\fmf{phantom}{o,j}
\fmf{plain,left}{m,n}
\fmf{plain,left}{n,o}
\fmf{plain,right,tension=0}{m,n}
\fmf{plain,right,tension=0}{n,o}
\fmf{plain,left,tension=0}{m,o}
\end{fmfgraph*}}
\ee
in the self energy; all other third order diagrams are automatically
included by self-consistency. This diagram now has one internal vertex whose
time index is not fixed by the two time indices that the self-energy
carries. It therefore gives a contribution to $\Sigma_{..}(t,t')$
which has an integral over this ``internal'' time; \eg\ for
$\Sigma_{22}$ we get a contribution
\be
\Sigma_{22}(t,t') \sim C(t,t')\int \!\!dt''\,
C(t,t'')C(t',t'')[R(t,t'')C(t',t'') + R(t',t'')C(t,t'')]
\ee
Diagrams of even higher order contain additional time integrals; in
general, then, the self-energy is a {\em functional} of the
response and correlation functions.  

An approximation, like that in (\ref{eq:MCTsigma12}) and (\ref{eq:MCTsigma22}), in which we keep only diagrams for $\Sigma(t,t')$ that are functions of $G(t,t')$ and $C(t,t')$  has come to be called a {\em mode coupling approximation}.  These have been studied extensively, often because they are exact for some mean-field (infinite-range) models \cite{bouchaud1996mode}.

\section{Ghosts and supersymmetry}
\label{StratonovichSUSY}

\subsection{Using Grassmann variables}

In Sec.~\ref{sec:perturbation}, we discussed how the results of the perturbation theory are independent of the value of $\lambda$ chosen in the discretization of the Langevin equation: the choice of $\lambda$ changes the Jacobian but does not affect correlation functions and other observables. As alluded to already in Sec.~\ref{MSR}, there is another way of including the Jacobian in the path integral using Grassmann variables, which is both conceptually interesting and can often simplify notation. This observation was first made, and its consequences on dynamics explored, by Feigel'man and Tsvelik \cite{feigel1982hidden,feigel1983hidden}, who followed earlier work by Parisi and Sourlas \cite{parisi1979random} on supersymmetric properties of an equilibrium system in a random external field. 
The current section is devoted to describing this approach.

Before getting into the details of the supersymmetric formalism, we need to familiarize ourselves with Grassmann variables. Grassmann variables are charaterized by the fact that they {\em anticommute} with each other.  Multiplication of them is also associative; i.e.\ for Grassmann variables $\xi_i$, we have
\bea
\xi_i (\xi_j \xi_k)= (\xi_i \xi_j) \xi_k \ \ \mbox{(associative)} \\
\xi_i \xi_j= - \xi_j \xi_i \ \ \mbox{(anti-commutative)}  
\eea
A consequence of anti-commutation is that 
\be
\xi_i^n=0\ \ \forall n>1 \label{eq:grass}
\ee
Grassmann variables can be added to each other and also multiplied by complex numbers: one says formally that they form an algebra, i.e.\ a vector field over the complex numbers endowed with a multiplication.  This means that, given\eq{eq:grass}, the most general functions of one and two Grassmann variables can be written respectively as
\bea
f(\xi_1) = c_0+ c_1 \xi_1   \\
f(\xi_1,\xi_2) = c_0+ c_1 \xi_1 + c_2 \xi_2 + c_{12} \xi_1 \xi_2,
\eea
where $c_0,c_1, c_2$ and $c_{12}$ are arbitrary complex numbers.

Integration and differentiation for Grassmann variables are defined by
\be
\frac{d}{d\xi_i} \xi_j = \delta_{ij}, \ \ \ \  \int d\xi = 0, \ \ \ \  \int d\xi\,\xi = 1,
\ee
and these lead to the following formulae that we will use later:
\bea
&\int d\xi (a + b \xi) = b,\ \ \ \ \int d\xi_1 d\xi_2  \,\xi_2 \xi_1=1,\\
&\frac{d}{d\xi} f(\xi)= \frac{d}{d\xi} (a+b \xi)= b, \ \ \ \ \frac{d}{d\xi_1} \xi_2 \xi_1 = - \xi_2.
\eea
As a consequence of the above, and using two independent sets of Grassmann variables $\xi_n$ and $\overline{\xi}_n$, one has the following important representation of the determinant of a matrix $\bm{A}$:
\be
|\bm{A}|=\int D[\xi\overline{\xi}] \exp\left \{ \sum_{mn} \overline{\xi}_m A_{mn} \xi_n \right \} ,
\label{eq:detAgrass}
\ee
where $D[\xi \overline{\xi}]=\prod_n d\xi_n d\overline{\xi}_n$. The integrand on the right hand side of \eref{eq:detAgrass} defines a Gaussian measure for Grassmann variables under which we have $\langle {\xi}_m \overline{\xi}_n \rangle = - (\bm{A}^{-1})_{mn}$.

Employing the representation \eref{eq:detAgrass} for the determinant of a matrix, we can write the Jacobian that appears in the transformation \eref{eq:jack} in the following way. First, we recall that the non-zero elements of the Jacobian are
\bea
\partial \eta_n/\partial \phi_{n+1}=1-\Delta \lambda f'_{n+1}\\
\partial \eta_n/\partial \phi_{n}=-1-\Delta (1-\lambda) f'_{n}
\eea
and therefore
\bea
\fl J(\phi) =  \int D[\xi\overline{\xi}] \exp\left \{\sum_n \overline{\xi}_n (1-\Delta \lambda f'_{n+1})\xi_{n+1}+ \sum_n \overline{\xi}_n [-1-\Delta (1-\lambda) f'_{n}] \xi_{n} \right \} 
\eea
where $\xi_n$ (with $n=1,\ldots,M$) and $\overline{\xi}_n$ (with $n=0,\ldots,M-1$) are Grassmann numbers.

The dynamical partition function is defined as before in (\ref{Z_av}) but with the average now also involving integration over the Grassmann variables,
\be
\langle \ldots \rangle_S = \int\! \frac{d\phi\,d\hph}{(2\pi)^M} \prod_n {d\xi_n d\overline{\xi}_n}\, \ldots \ \exp(-S) \equiv \int D[\phi\hat\phi] D[\xi\overline\xi]\, \ldots\ \exp(-S)
\label{av_def_Grassmann}
\ee
The action reads
\bea
S &=& \sum_n 
\left(T\Delta\hph_n^2 - i\hph_n 
\left\{-\phi_{n+1}+\phi_n + \Delta[(1-\lambda)f_n +
\lambda f_{n+1}]\right\}\right)\nonumber
\\
&&{}-\sum_n \overline{\xi}_n (1-\Delta \lambda f'_{n+1})\xi_{n+1}- \sum_n \overline{\xi}_n (-1-\Delta (1-\lambda) f'_{n}) \xi_{n} 
\label{action_Grassmann}
\eea
where the second line replaces the last term in\eq{action}.
In the continuous time limit $\Delta\to 0$ this can be written as
\bea
S&=&{}\int dt \left\{ T \hat{\phi}^2 +i \hat{\phi}[\partial_t \phi-f(\phi)]\right\}-\int dt\,\overline{\xi} [ \partial_t -f'(\phi) ] \xi    \label{eq:actionwithghosts}    
\eea

Let us see how the inclusion of the Grassmann ``ghosts'' works out for our example case of $f(\phi)= -\mu \phi - (g/3!) \phi^3$. Going back to discretized time temporarily will help us understand how to treat equal-time correlations. With $f(\phi)$ as given, the action can be written as
\bea
S&=&S_0+\Si\\
S_0 &=& \sum_n
\left(T\Delta\hph_n^2 - i\hph_n 
\left\{-\phi_{n+1}+\phi_n - \mu\Delta[(1-\lambda)\phi_n +
\lambda \phi_{n+1}]\right\}\right)
\nonumber
\\
&&{}-\sum_n \left\{\overline{\xi}_n (1+\Delta \lambda \mu)\xi_{n+1}+
\overline{\xi}_n [-1+\Delta (1-\lambda)\mu] \xi_{n}\right\}
\\
\Si &=& 
i\frac{g}{3!}\Delta \sum_n\hph_n 
[(1-\lambda)\phi_n^3 +
\lambda \phi_{n+1}^3]
\nonumber
\\
&&{}-\Delta \frac{g}{2} \sum_n \left[
\overline{\xi}_n \lambda \phi_{n+1}^2\xi_{n}+ \overline{\xi}_n (1-\lambda) \phi_n^2 \xi_{n}\right] 
\eea
The coefficient matrix $\bm{A}$ appearing in the ghost term of the
bare action $S_0$
has entries $1+\Delta\lambda\mu$ on the main diagonal and
$-1+\Delta(1-\lambda)\mu$ on the diagonal below. This matrix is easily
inverted to show that the ghost covariance is
\be
\langle \xi_m \overline{\xi}_n\rangle_0 = 
-\frac{1}{1+\Delta\lambda\mu}\left(\frac{1-\Delta(1-\lambda)\mu}
{1+\Delta\lambda\mu}\right)^{m-n-1}= -\exp[-\mu(m-n-1)\Delta]
\label{xi_correlator_bare}
\ee
for $m>n$ and 0 otherwise; the last expression applies for
$\Delta \to 0$. The ghost correlator is therefore causal and reads in
the continuum limit:
\be
\langle {\xi}(t) \overline{\xi}(t')\rangle_0 = -\Theta(t-t')\exp [-\mu (t-t')].	\label{eq:ghostcorr}
\ee

While this is $\lambda$-independent, the dependence on $\lambda$
reappears in how equal-time Wick contractions are treated in the
perturbative expansion in powers of $-\Si$. To see this, note that up
to vanishing corrections the interacting action is
\bea
\Si &=& 
i\frac{g}{3!}\Delta \sum_n [\lambda\hph_{n-1}+(1-\lambda)\hph_n]\phi_n^3
\nonumber
\\
&&{}
-\Delta \frac{g}{2} \sum_n [
\lambda \overline{\xi}_{n-1} + (1-\lambda) \overline{\xi}_{n}] \phi_{n}^2 \
\xi_{n} 
\eea
The square brackets in the first line, including the factor $i$, define what we previously called
the response field $\eta_{2n}$, which has equal-time correlator with
$\phi_n$ of $R_{nn}^0=\lambda$. Similarly we could now define the
combination in square brackets in the second line as a new Grassmann
response field $\tilde\xi_n$, which has equal-time correlator with
$\xi$ of $\langle \xi_n \tilde\xi_n \rangle_0 = \lambda \langle \xi_n
\overline{\xi}_{n-1}\rangle_0 = -\lambda$ using\eq{xi_correlator_bare}.
If for simplicity of notation we do not distinguish between
$\tilde\xi$ and $\overline{\xi}$ (and similarly $\eta_2$ and
$\hat\phi$) then the upshot of this discussion is that for $\Delta\to 0$ one can work
directly with the continuous-time version
\be
\Si = \frac{g}{3!} \int dt  (i\hat{\phi} \phi^3-3\overline{\xi} \phi^2 \xi)
\ee
of the interacting action, provided that one remembers that the
equal-time ghost correlator has to be set to $\langle
\xi(t)\overline{\xi}(t)\rangle_0=-\lambda$ in any Wick contractions, and similarly
$\langle \phi(t)\,i\hat\phi(t)\rangle_0=\lambda$. Note that there are never any contractions between ordinary and Grassmann variables because the average of any single Grassmann variable over a (Grassmann) Gaussian vanishes.

To see how the perturbation theory with ghosts works in practice, consider the response function to first order in the perturbation:
\bea
 \langle \phi(t) \hat{\phi}(t') \rangle_S&=&\langle \phi(t) \phi(t') \exp[-\Si]\rangle_0
=\langle \phi(t) \hat{\phi}(t') [1-\Si] \rangle_0 + {\cal O}(g^2)\\
&=&\langle \phi(t) \hat{\phi}(t') \rangle_0-\frac{ig}{3!} \int dt'' \langle \phi(t)\hat{\phi}(t') \hat{\phi}(t'')\phi^3(t'') \rangle_0\nonumber\\
&&{}+\frac{g}{2} \int dt''  \langle \phi(t)\hat{\phi}(t') \phi^2(t'') \bar{\xi}(t'')\xi(t'') \rangle_0 + {\cal O}(g^2)
\label{response_ghost_perturb}
\eea
The physical piece of this is the first term and the contraction without equal-time response factors in the first integral:
\bea
 \langle \phi(t) \hat{\phi}(t') \rangle
&=&\langle \phi(t) \hat{\phi}(t') \rangle_0-\frac{ig}{2} \int dt'' \langle \phi(t)\hat{\phi}(t'')\rangle_0 \langle \phi(t'')\hat{\phi}(t')\rangle_0 \langle \phi^2(t'')\rangle_0
\eea
When $\lambda\neq0$, i.e.\ for any convention other than Ito, there are two other nonzero contractions of the first integral:
\bea
&&{}-\frac{g}{2}  \langle \phi(t)\hat{\phi}(t')  \rangle_0 \int dt'' \langle \phi^2(t'') \rangle_0 \langle \phi(t'')i\hat{\phi}(t'') \rangle_0 \nonumber \\
&&\ \ \ \ {}-g \int dt'' \langle \phi(t) \phi(t'') \rangle_0 \langle \phi(t'') \hat{\phi}(t') \rangle_0  \langle  \phi(t'')i\hat{\phi}(t'') \rangle_0
\label{extra_terms}
\eea
In addition the two possible Wick contractions of the ghost term in the second line of\eq{response_ghost_perturb} give, bearing in mind that $\langle \overline{\xi}\xi\rangle_0 = -\langle \xi\overline{\xi}\rangle_0$,
\bea
&&{}-\frac{g}{2}  \langle \phi(t)\hat{\phi}(t')  \rangle_0 \int dt'' \langle \phi^2(t'') \rangle_0 \langle \xi(t'') \ovr{\xi}(t'') \rangle_0\nonumber \\
&&\ \ \ \  {}-g   \int dt'' \langle \phi(t) \phi(t'') \rangle_0 \langle \phi(t'')\hat{\phi}(t') \rangle_0  \langle \xi(t'')\ovr{\xi}(t'')\rangle_0 
\label{extra_terms2}
\eea
Because $\langle \phi \,i\hat\phi\rangle_0=\lambda$ at equal-time while the ghost correlator has the {\em opposite} sign $\langle \xi\ovr{\xi}\rangle_0 = -\lambda$, the terms in\eq{extra_terms} and\eq{extra_terms2} exactly cancel each other as they should.

In other words, whether or not we include the Jacobian, the extra terms\eq{extra_terms} do not appear in the perturbation theory, either because they are simply equal to zero, or because they cancel with the additional terms\eq{extra_terms2} that arise from the ghost correlation functions from the Grassmann representation of the Jacobian. As will be elaborated in the next section, this is formally a consequence of a symmetry of the theory that represents the fact that the path probabilities are normalized to one. 

\subsection{Manifestly supersymmetric description}

Let us now define the {\em superfield} $\Phi$ as
\be
\Phi = \phi+\overline{\theta} \xi+\overline{\xi}\theta +i \overline{\theta} \theta \hat{\phi}  \label{eq:superfield}
\ee
where $\theta$ and $\overline{\theta}$ are themselves Grassmann variables and are sometimes referred to as Grassmann time.   The action\eq{eq:actionwithghosts} can then be written in terms of $\Phi$ in a compact way that reveals unexpected symmetries of the problem.

To see this, it is convenient to define
\be
V(\phi) = - \int^{\phi} d\phi' f(\phi'),
\ee
so $f(\phi)= -dV(\phi)/d\phi$.
The existence of the potential $V$ is crucial for the supersymmetric description; it cannot be used for systems of more than one variable when these are non-equilibrium in the sense that the drift $f$ cannot be written as a gradient.

In our current one-dimensional example where a potential can always be defined we have, by Taylor expansion around $V(\phi)$ and throwing away terms that vanish because of~(\ref{eq:grass}),
\bea
V(\Phi) &=& V(\phi)-(\overline{\theta} \xi+\overline{\xi}\theta+i \overline{\theta} \theta \hat{\phi}) f(\phi)-\frac{1}{2}(\overline{\theta} \xi+\overline{\xi}\theta+i \overline{\theta} \theta \hat{\phi})^2 f'(\phi) \\
&=& V(\phi)-(\overline{\theta} \xi+\overline{\xi}\theta+i \overline{\theta} \theta \hat{\phi}) f(\phi)
+ \overline{\theta}\theta \overline{\xi}\xi f'(\phi),
\eea
which leads to 
\be
\int d\theta d\overline{\theta} \,V(\Phi) = -i \hat{\phi} f(\phi)+\overline{\xi} f'(\phi)\xi,
\ee
giving us two of the terms in\eq{eq:actionwithghosts}.  

The remaining terms can be written in terms of derivatives of $\Phi$.  We have
\bea
\partial_{\theta} \Phi &=& - \overline{\xi} - i \overline{\theta}\hat{\phi}			\label{eq:ptheta}  			\\
\partial_{\overline{\theta}} \Phi &=& \xi + i\theta \hat{\phi}   \label{eq:pthetabar}  			\\
\theta \partial_t \Phi  &=&  \theta \partial_t \phi+ \theta \overline{\theta} \partial_t \xi.  \label{eq:pt}
\eea 
If we then evaluate the quantity
\be
T \int  d\theta d{\overline{\theta}} (\partial_{\theta} \Phi )(\partial_{\overline{\theta}} \Phi),
\ee
we find that we get the $T\hat{\phi}^2$ term in the action\eq{eq:actionwithghosts}. Similarly, we find
\be
\int  d\theta d{\overline{\theta}} (\partial_{\theta} \Phi) (\theta \partial_{t} \Phi )= -i\hat{\phi}\partial_t\phi +\overline{\xi}\partial_t{\xi},
\ee
which are the negatives of the terms in\eq{eq:actionwithghosts} involving time derivatives.  
Putting all these results together and defining $d\tau\equiv dt\, d\theta\, d\overline{\theta}$, we can write the action in the form
 \be
S= \int d\tau  \left \{ \partial_{\theta} \Phi [T \partial_{\overline{\theta}}\Phi - \theta \partial_t \Phi] + V(\Phi) \right \}. 											\label{eq:dynSSaction}
\ee
It will be handy to introduce the notation
\bea
{\rm D} &=&  T\partial_{\overline{\theta}} - \theta \partial_t,				\\
\overline{\rm D} &=& \partial_\theta						\label{eq:DandDbar}
\eea
so that
\be
S= \int d\tau  \left [\overline{\rm D} \Phi  {\rm D}\Phi+ V(\Phi) \right ]. 	\label{eq:dynSSaction2}	 
\ee

Up to here the formalism is general enough -- subject to the existence of the potential $V$ -- that it can describe also non-stationary dynamics, e.g.\ relaxation to equilibrium. From now on we 
restrict ourselves further by assuming we have a stationary  state. The supersymmetric action then has several symmetries; the obvious one is time translation invariance.  But it is also invariant under other ``translations'' that involve shifts in the Grassmann times $\theta$ and $\overline{\theta}$.  In what follows, we identify these invariances and investigate their physical meanings 
\cite{zinn2002quantum,kurchan1992supersymmetry}. 

Consider a ``translation'' generated by  the operator 
\be
{\rm D}' \equiv \partial_{\overline{\theta}}.
\ee
This produces a shift $\overline{\theta}  \to \overline{\theta} + \epsilon$ and a change in $\Phi$:
\be 
\Phi \to \Phi + \epsilon \partial_{\overline{\theta}}\Phi  ,
\ee
where $\epsilon$ is a Grassmann ``infinitesimal" that acts like a separate Grassmann variable. Now,  by using \eref{eq:pthetabar} (or simply by substituting $\overline{\theta}  \to \overline{\theta} + \epsilon$ in  the definition \eref{eq:superfield} of the superfield), we find
\be
\Phi \to \Phi + \epsilon \xi + i \epsilon \theta \hat{\phi} =
(\phi+ \epsilon \xi) +\overline{\theta}\xi +(\overline{\xi} +i \epsilon \hat{\phi})\theta 
+ i \overline{\theta} \theta \hat{\phi}.      \label{eq:epsilonPhi}
\ee
But according to the definition of the superfield, whatever appears in the expression \eref{eq:epsilonPhi} not mutiplied by $\theta$, $\overline{\theta}$ or $\overline{\theta}\theta$ should be identified as the new $\phi$, and whatever appears multiplied (from the right) by $\theta$ should be identified as the new $\overline{\xi}$. The component fields therefore transform as
\bea
\phi &\to& \phi + \epsilon \xi 			\label{eq:dphi}	  \\
\xi &\to& \xi						\label{eq:dxi}				  \\
\overline{\xi }&\to& \overline{\xi} + i\epsilon \hat{\phi},    \label{eq:dxibar}     \\
\hat{\phi} &\to& \hat{\phi}.  			\label{eq:dphihat}
\eea
This transformation is called supersymmetric because it mixes ``bosonic" degrees of freedom (i.e.\ $\phi$ and $\hat{\phi}$) with ``fermionic" (i.e.\ Grassmann) ones.
  
To verify that such a transformation is indeed a symmetry of the model, we proceed analogously to what we would do to test, say, rotational invariance in an ordinary field theory with vector fields.  We start by performing the ``rotation'' (\ref{eq:dphi}--\ref{eq:dphihat}) on the superfield components in each term of the integrand in \eref{eq:dynSSaction2}, i.e.\ we substitute the transformed $\Phi$ and carry out the required derivatives with respect to time and Grassmann time. In general, this leads to a change in the integrand, which we then integrate over $\tau$ (i.e.\ over $\theta$, $\overline{\theta}$ and $t$). If the result is zero we have a symmetry. 
As an example, let us see how the shift generated by ${\rm D}'$ affects the ``kinetic'' term $\overline{\rm D}\Phi {\rm D}\Phi$ in the action.  
From \eref{eq:ptheta} we see, using \eref{eq:dxibar} and \eref{eq:dphihat}, that
\be
\overline{\rm D}\Phi \to \overline{\rm D}\Phi  - i \epsilon \hat{\phi}.
\ee
For the change in ${\rm D}\Phi$, we see from (\ref{eq:pthetabar}) that the first term $\partial_{\overline{\theta}}\Phi$ does not change as it involves only $\xi$ and $\hat{\phi}$, and these do not change under (\ref{eq:dxi},\ref{eq:dphihat}).  From \eref{eq:pt}, \eref{eq:dphi} and \eref{eq:dxi}, the change in $\theta \partial_t \Phi$ is
\be
 \theta \partial_t \Phi \to \theta \partial_t \Phi - \epsilon \theta \dot{\xi}.
 \ee
Putting these contributions together, the change in $\overline{\rm D}\Phi {\rm D}\Phi$  is
\be
(-\overline{\xi} - i\overline{\theta} \hat{\phi})\epsilon \theta \dot{\xi}
- i \epsilon \hat{\phi} [T(\xi + i \theta \hat{\phi}) - \theta \dot{\phi} - \theta \overline{\theta} \dot{\xi}].
\ee
Only terms proportional to $\theta \overline{\theta}$ will survive the integration over $\theta$ and $\overline{\theta}$, but these cancel:
\be
-i\epsilon \theta \overline{\theta}  \hat{\phi} \dot{\xi} 
+ i \epsilon \theta \overline{\theta} \hat{\phi} \dot{\xi} = 0.
\ee
A similar calculation shows that the ``potential'' term $\int d\tau\, V(\Phi)$ is also unchanged.  Thus, the action $S$ is invariant under ${\rm D}' = \partial_{\overline{\theta}}$.

An analogous calculation shows that the kinetic term is {\em{not}} invariant under shifts generated by $\overline{\rm D}=\partial_{\theta}$. However, we can try combining a shift in $\theta$ with one in time, using the generator
\be
\overline{\rm D}' = \partial_{\theta} +\alpha \overline{\theta} \partial_t							\label{eq:combinedshift}
\ee
and see whether there is a value of $\alpha$ for which $\overline{\rm D}\Phi {\rm D}\Phi$ is invariant.  Now the transformation of $\xi$ acquires a new term proportional to $\partial_t \phi$,
\be
\xi \to \xi + \epsilon (i\hat{\phi} -\alpha \partial_t \phi),					\label{eq:newdxi}
\ee
and $\hat{\phi}$ is also changed, proportional to $\partial_t\overline{\xi}$:
\be
\hat{\phi} \to \hat{\phi} - i(\alpha \partial_t\overline{\xi})\epsilon\ .			\label{eq:newdphihat}
\ee 
The remaining variables $\phi$ and $\overline{\xi}$ transform according to
\bea
\phi &\to& \phi + \overline{\xi} \epsilon  			\label{eq:newdphi}  \\
\overline{\xi} &\to& \overline{\xi}, 				\label{eq:newdxibar}
\eea
Then, after some algebra, we find that, under our trial $\overline{\rm D}'$, $\overline{\rm D}\Phi {\rm D}\Phi$ is changed by
\bea 
&&{}-\overline{\xi} [(1 - \beta \theta \overline{\theta}\partial_t)(i\hat{\phi} -\alpha \dot{\phi})
+(\alpha - \beta)\theta \partial_t\overline{\xi}] \epsilon
-i \overline{\theta} \hat{\phi} (i\hat{\phi} - \alpha \dot{\phi})\epsilon  \nonumber \\
&&{}-i(\alpha - \beta)\hat{\phi}\overline{\theta} \theta \partial_t \overline{\xi}\epsilon
- \alpha \overline{\theta} (\partial_t \overline{\xi}) \epsilon [\xi +\theta (i\hat{\phi} - \beta \dot{\phi})]
\eea
(here $\beta = 1/T$).  Integrating over $\theta$ and $\overline{\theta}$ leaves
\be
- \beta \overline{\xi} \partial_t(i\hat{\phi} - \alpha \dot{\phi})\epsilon 
- \alpha (\partial_t \overline{\xi}) \epsilon (i\hat{\phi} - \beta \dot{\phi}) 
- i(\alpha - \beta)\hat{\phi}\partial_t\overline{\xi}\epsilon,
\ee
so we see that with the choice $\alpha = \beta$ this just reduces to
\be 
- \beta \partial_t[\overline{\xi}(i\hat{\phi} - \dot{\phi})]\epsilon\ .
\ee
This vanishes on integration over $t$ (for stationary initial and final states), proving the invariance of this part of the action. Again the proof of invariance for the $V(\Phi)$ term is similar to that for ${\rm D}'$, so the  total action is invariant under $\overline{\rm D}'$.

The reader who is uneasy with the formal manipulations using superfields here can check these results by applying the transformations (\ref{eq:dphi}--\ref{eq:dphihat}) for ${\rm D}'$ and (\ref{eq:newdxi}--\ref{eq:newdxibar}) for $\overline{\rm D}'$ directly to $\phi$, $\hat{\phi}$, $\xi$ and $\overline{\xi}$, in the form (\ref{eq:actionwithghosts}) of the action not using superfields.

To see the meaning of these supersymmetries, we consider the supercorrelation function
\be
Q(1,2) = \langle \Phi(1) \Phi(2) \rangle,
\ee
where $1$ stands for $(t_1, \theta_1, \overline{\theta_1})$ and analogously for $2$. When we use (\ref{eq:superfield}) and expand the product,  many terms vanish, either because the averages are of products of Grassmann variables with ordinary ones or of a pair of $\hat{\phi}$'s.  The remaining terms are
\be
Q(1,2) = \langle \phi_1 \phi_2 \rangle 
+ \overline{\theta}_1 \theta_2 \langle \xi_1 \overline{\xi}_2 \rangle
+ \theta_1 \overline{\theta}_2  \langle \overline{\xi}_1 \xi_2 \rangle 
+ i\overline{\theta}_1 \theta_1 \langle \hat{\phi}_1 \phi_2 \rangle
+ i\overline{\theta}_2 \theta_2 \langle \phi_1 \hat{\phi}_2 \rangle,     \label{eq:Qred}
\ee
where $\phi_1$ means $\phi(t_1)$, etc.  

From the invariance of the action under ${\rm D}'$ (translations in $\overline{\theta}$), we have 
\bea
{\rm D}'Q(1,2)  &=& ({\rm D}'_1 + {\rm D}'_2) Q(1,2)		
= (\partial_{\overline{\theta}_1}+\partial_{\overline{\theta}_2})Q(1,2)   		\nonumber \\
&=& \theta_2 (\langle \xi_1 \overline{\xi}_2 \rangle +  i \langle \phi_1 \hat{\phi}_2 \rangle)
+  \theta_1 (-\langle \overline{\xi}_1 \xi_2 \rangle  + i \langle \hat{\phi}_1 \phi_2 \rangle  ) = 0.
\label{eq:Dprimeconstraint}
\eea
The vanishing of the term proportional to $\theta_2$ says that the ghost correlation function $\langle \xi_1 \overline{\xi}_2 \rangle$ has to be the negative of the response function  $ i \langle \phi_1 \hat{\phi}_2 \rangle$ (as in (\ref{eq:ghostcorr})).  The vanishing of the term proportional to $\theta_1$ says the same thing if we notice that the ghost correlation function here is $\langle \overline{\xi} \xi \rangle$, not $\langle \xi \overline{\xi} \rangle$. Thus, invariance under ${\rm D}'$, through its enforcement of the cancellation of disconnected diagrams, is just conservation of probability.

Analogously, for the $\overline{\rm D}'$ symmetry (\ref{eq:combinedshift}), we have 
\be
\overline{\rm D}'Q(1,2) = (\overline{\rm D}'_1+\overline{\rm D}'_2)Q(1,2) = (\partial_{\theta_1}+\partial_{\theta_2}+ \beta \overline{\theta}_1 \partial_{t_1}
+ \beta \overline{\theta}_2 \partial_{t_2})Q(1,2) = 0.
\ee
This gives
\be
(\overline{\theta}_1 - \overline{\theta}_2)
(\langle \phi_1 \hat{\phi}_2 \rangle - \langle \hat{\phi}_1 \phi_2 \rangle )
+ \beta \overline{\theta}_1  \partial_{t_1} \langle \phi_1 \phi_2 \rangle
+ \beta \overline{\theta}_2  \partial_{t_2} \langle \phi_1 \phi_2 \rangle = 0.
\ee
For $t_1 > t_2$, $\langle \hat{\phi}_1 \phi_2 \rangle$ vanishes, so we have, also using $\partial_{t_2} \langle \phi_1 \phi_2 \rangle = -\partial_{t_1} \langle \phi_1 \phi_2 \rangle$ from time translation invariance,
\be
(\overline{\theta}_1 - \overline{\theta}_2)
(\langle \phi_1 \hat{\phi}_2 \rangle
 + \beta  \partial_{t_1} \langle \phi_1 \phi_2 \rangle) = 0.
 \ee
Thus, 
\be
\langle \phi_1 \hat{\phi}_2 \rangle = -  \beta \partial_{t_1} \langle \phi_1 \phi_2 \rangle,
\ee
which is the fluctuation-dissipation theorem. 

To summarize, the theory has three invariances: time translation, ${\rm D}'$ and $\overline{\rm D}'$. ${\rm D}'$ expresses conservation of probability, and $\overline{\rm D}'$ expresses the fluctuation-dissipation theorem, i.e.\ the fact that the system is in equilibrium. 

So far, we have treated a single-site problem.  It is straightforward to extend the formalism to multiple degrees of freedom $\phi_i$. 
However, as emphasized before, the supersymmetric construction is permitted only when the drift is the negative gradient of a potential, $f_i = -\partial V/\partial \phi_i$.  Otherwise the $\overline{\rm D}'$ supersymmetry fails.  This means that the fluctuation-dissipation theorem is not obeyed; the system, even though it may possess a steady state, is not in equilibrium. (Of course, it is well-known from simple arguments making no reference to supersymmetry that models with non-gradient drifts, as arising e.g.\ from asymmetric coupling matrices, do not satisfy the fluctuation-dissipation theorem.)

\subsection{Superdiagrams}

In addition to the insight it provides into the symmetries of the problem, the supersymmetric formulation can also be of practical advantage in calculations.  For example, in diagrammatic perturbation theory, one needs only draw the diagrams for the static problem.   Another example can be found in Biroli's analysis \cite{biroli1999dynamical} of dynamical TAP equations for the $p$-spin glass, where the entire structure of the argument and the equations could be carried over from the static treatment of Plefka \cite{plefka1982convergence}. Here we sketch how to do diagrammatic perturbation theory in the superfield language and show, in a simple example, how it reduces to the diagrams we had in the MSR formalism with the $\langle \phi (t) \phi (t')\rangle$ and $\langle \phi (t) \hat{\phi} (t')\rangle$ correlation functions \cite{bouchaud1996mode}.

We write our standard model\eq{eq:one_body_model} in the form
\be
S = S_0 + \Si,
\ee
with
\be
S_0 = \int d\tau \left ( \overline{\rm D}\Phi {\rm D}\Phi + \lhalf \mu \Phi^2\right),	\label{eq:Szero}
\ee
and
\be
\Si = \frac{g}{4!} \int d\tau\, \Phi^4\ .
%(\tau).
\ee
To do perturbation theory, we can expand $\exp (-\Si)$ in $g$ and apply Wick's theorem to evaluate the resulting averages, just as we did in Sec.~\ref{Diags}. 
Wick's theorem holds also for Grassmann variables although in principle one has to be careful with sign changes that arise from changing the order of the variables when performing contractions. 
For example, $\langle \xi_1 \overline{\xi}_1 \xi_2 \overline{\xi}_3\rangle = 
\langle \xi_1 \ovr{\xi}_1\rangle 
\langle \xi_2 \ovr{\xi}_3 \rangle
- \langle \xi_1 \ovr{\xi}_3\rangle 
\langle \xi_2 \ovr{\xi}_1\rangle$.
Fortunately this is not an issue in our context as we only need averages of powers of the superfield $\Phi$, and from\eq{eq:superfield} $\Phi$ only contains products of {\em pairs} of Grassmann variables:  commuting such pairs through each other never gives any minus signs.

The first thing we need for the perturbation theory is the correlation function or propagator of the noninteracting system.  Integrating (\ref{eq:Szero}) by parts, we get
\be
S_0 = \int d\tau \left ( -\Phi \overline{\rm D} {\rm D}\Phi + \lhalf \mu \Phi^2\right).		\label{eq:Soo}
\ee
It is convenient to write this as
\be
S_0 = \lhalf \int d\tau \left [(- \Phi (  [\overline{\rm D}, {\rm D}]_-  +  [\overline{\rm D}, {\rm D}]_+) \Phi 
+ \mu \Phi^2\right),											\label{eq:Scac}
\ee
where $[\ldots]_-$ and $[\ldots]_+$ denote the commutator and anti-commutator respectively. It is then simple to show that
\be
[\overline{\rm D}, {\rm D}]_+ = - \partial_t
\ee
and 
\be
 [\overline{\rm D}, {\rm D}]_- = 2 T\partial_\theta \partial_{\overline{\theta}} + 2\theta \partial_\theta \partial_t - \partial_t.									\label{eq:DbarDcomm}
\ee
The term in (\ref{eq:Scac}) involving the anticommutator can be neglected because it vanishes on time integration, so we can write $S_0$ in the appealing form
\be
S_0 = \lhalf \int d\tau \Phi ( - {\rm D}^{(2)}+ \mu) \Phi,							 
\ee
with\footnote{We could of course add any term proportional to $\partial_t$ to ${\rm D}^{(2)}$ without changing the action, i.e.\ we could put {\em any} coefficient $a$ in front of the $\partial_t$ term in (\ref{eq:DbarDcomm}) and the action would remain unchanged.  Choosing $a=2$ would correspond to including the anticommutator term from (\ref{eq:Scac}). However, the choice $a=1$, i.e.\ ${\rm D}^{(2)} = [\overline{\rm D}, {\rm D}]_-$, will prove convenient when we want to invert this operator to find the unperturbed superfield correlation function.}  
\be
{\rm D}^{(2)} \equiv[\overline{\rm D}, {\rm D}]_-\ .
\ee 
From this, we identify
\be
Q_0^{-1} = -{\rm D}^{(2)} + \mu 
\ee
as the inverse of the free propagator:
\be
(-{\rm D}^{(2)} + \mu)Q_0(1,2) = \delta(1,2) 
\equiv \delta(t_1-t_2)(\overline{\theta}_1 -\overline{\theta}_2)(\theta_1-\theta_2),
\ee
where we have used the fact that $(\overline{\theta}_1 -\overline{\theta}_2)(\theta_1-\theta_2)$ acts as a delta-function in the Grassmann times.  Now, multiplying by $({\rm D}^{(2)} + \mu)$ from the left, using the fact that $({\rm D}^{(2)})^2 = \partial^2_{t}$, and Fourier transforming in time, we arrive at 
\be
\fl Q_0 (\theta_1, \overline{\theta}_1, \theta_2, \overline{\theta}_2; \omega)
= \frac{1}{\omega^2 + \mu^2}[2T + (\overline{\theta}_1 -\overline{\theta}_2)(\theta_1-\theta_2)\mu 
+ (\overline{\theta}_1 +\overline{\theta}_2)(\theta_1+\theta_2)i\omega].					
\ee
Back in the time domain, this is
\be
\fl Q_0(1,2) = \frac{T}{\mu} \exp (-\mu |t|) 
+  (\overline{\theta}_2 -\overline{\theta}_1)\theta_2 \exp (-\mu t)\Theta(t)
+  (\overline{\theta}_1 -\overline{\theta}_2)\theta_1 \exp (\mu t)\Theta(-t).			\label{eq:Gzero}
\ee	
where $t=t_1-t_2$. 
It is comforting to confirm that we can get this result in another, simpler way, using the representation (\ref{eq:Qred}) with the equalities implied by ${\rm D}'$ invariance (\ref{eq:Dprimeconstraint}):
\be
Q(1,2) = \langle \phi_1 \phi_2 \rangle 
+ (\overline{\theta}_2 -\overline{\theta}_1) (\theta_2 \langle \phi_1 i\hat{\phi}_2 \rangle 
- \theta_1 \langle i\hat{\phi_1}  \phi_2 \rangle ).							\label{eq:Qform}
\ee
Using the free correlation and response functions (\ref{eq:freecorr}) and (\ref{response}) here then leads to (\ref{eq:Gzero}).

To see how the diagrammatics work in this formalism, consider the second-order watermelon graph\eq{eq:watermelon} for the self-energy,   
\be
\Sigma(1,2) = \left(-\frac{g}{4!}\right)^2 \times 4 \times 4 \times 3! \times  Q(1,2)^3 =  \frac{g^2}{3!} 		Q(1,2)^3			\label{eq:melon}
\ee
(Here we are doing the resummed expansion in which $\Sigma$ is a functional of the full correlation function $Q$, not just $Q_0$, as in Sec.~\ref{sec:example}.) 

From the representation\eq{eq:Qform}, the part of $Q(1,2)$ with no Grassmann times multiplying it is the correlation function, and the parts mutiplied by $\pm (\overline{\theta}_2 -\overline{\theta}_1) \theta_{1,2}$ are the retarded and advanced response functions, respectively.   Expanding $Q(1,2)^3$, we get
\be
\Sigma(1,2)  = \frac{g^2}{3!} [\langle \phi_1 \phi_2 \rangle^3 + 
3 \langle \phi_1 \phi_2 \rangle^2(\overline{\theta}_2 -\overline{\theta}_1) (\theta_2  \langle \phi_1 i\hat{\phi}_2 \rangle - \theta_1 \langle  i\hat{\phi}_1 \phi_2 \rangle) ].			\label{eq:SSSigma}
\ee
This has the same form as (\ref{eq:Qform}). We note that the components of $\Sigma(1,2)$ here are just the $\mathcal{O}(g^2)$ contributions to the self-energies $\Sigma_{12}$, $\Sigma_{21}$ and $\Sigma_{22}$ 
that we found in the conventional MSR theory (\ref{eq:MCTsigma12})  and (\ref{eq:MCTsigma22}), including the factor of 3 in $\Sigma_{12}$ and $\Sigma_{21}$. 

It is useful to discuss in more detail how 
perturbation-theoretic corrections to $Q(1,2)$ in the supersymmetric formulation correspond to results obtained in the conventional MSR theory.  The two theories differ superficially in two ways: (1) In the conventional theory we have to keep track of two kinds of correlations functions ($\langle \phi \phi \rangle$ and $i\langle \phi \hat{\phi}\rangle$) while in the supersymmetric formulation we have only one (super)field and need only draw the diagrams we would have in statics. (2) In the supersymmetric theory both real and Grassmann times of intermediate vertices in the graphs are integrated over, while in the conventional theory only ordinary times are integrated over.  To compare the two ways of doing the calculation, we consider the first term obtained in expanding the Dyson equation in $\Sigma(1,2)$:
\be
\Delta Q(1,4) = \int d\tau_2 d\tau_3\, Q(1,2) \Sigma(2,3) Q(3,4),
\ee 

To resolve $\Delta Q(1,4)$ into components, we use first the fact that Grassmann factors of the form $(\overline{\theta}_2 -\overline{\theta}_1) \theta_2$ are idempotent under convolution, e.g.
\be
\int d\theta_2 d\overline{\theta}_2\, (\overline{\theta}_2 -\overline{\theta}_1) \theta_2
(\overline{\theta}_3 -\overline{\theta}_2) \theta_3 =  (\overline{\theta}_3 -\overline{\theta}_1) \theta_3.
\ee
That means that the retarded part of $\Delta Q(1,4)$ (the part proportional to $ (\overline{\theta}_4 -\overline{\theta}_1) \theta_4$) is, in the notation we have used earlier ($\langle \phi (t_1)\phi(t_2) \rangle = C(t_1,t_2)$, $i \langle \phi(t_1)\hat{\phi}(t_2)\rangle = R(t_1,t_2)$) 
\be
\int dt_2 dt_3\, R(t_1,t_2) \left[ \frac{g^2}{3!}\,3 (C(t_2,t_3))^2 R(t_2,t_3) \right]  R(t_3,t_4),	\label{eq:retpart}
\ee
and analogously for the advanced part.
 
We also get a contribution to $\Delta Q(1,4)$ from the first term in $\Sigma(2,3)$.  Since that term contains no Grassmann times, this contribution will contain a factor 
\be
\int d\theta_2 d\overline{\theta}_2 \,Q(1,2)\cdot \int d\theta_3 d\overline{\theta}_3 \,Q(3,4). 
\ee
These integrations pick out factors of the retarded function $R(t_1,t_2)$ and the advanced function $R(t_4,t_3)$, respectively, so we find a contribution, involving no Grassmann times, of 
\be
\int dt_2 dt_3\, R(t_1,t_2) \left[ \frac{g^2}{3!} (C(t_2,t_3))^3 \right]  R(t_4,t_3). \label{eq:cpart}
\ee

These are exactly the second-order contributions to $R(t_1,t_4)$ and $C(t_1,t_4)$ that we would find in the conventional formulation from expanding the Dyson equation to first order in the self-energies $\Sigma_{12}$ and $\Sigma_{22}$ in (\ref{eq:MCTsigma12}) and (\ref{eq:MCTsigma22}), again up to $\mathcal{O}(g)$ terms not written explicitly here. Thus, because of the algebra of the Grassmann times in the supersymmetric formulation, the results of the multiplication and convolution of the supercorrelation functions, when reduced to components, reproduce the terms found in the conventional MSR theory.  This result extends to all graphs in perturbation theory, because it depends only on (1) the idempotency of factors like $(\overline{\theta}_2 -\overline{\theta}_1) \theta_2$ and (2) the fact that multiple correlator lines between a pair of interaction vertices, like those in (\ref{eq:melon}), combine as in (\ref{eq:SSSigma}). 

\section{An interacting example}
To generalize the discussion so far to more interesting interacting models one can for example add a linear interaction between different degrees of freedom or ``soft spins" $\phi_i$. This gives the Langevin equation of motion
\be
\partial_t \phi_i =
%- \partial_{\phi_i}  H[\phi] + \zeta_i(t)= 
- \mu \phi_i - \frac{g}{3!} \phi^3_i+\sum_j J_{ij} \phi_j +h_i + \zeta_i(t)
\label{eq:phiint}
\ee
where we assume that there are no self-interactions, hence $J_{ii}=0$.
If the couplings are otherwise symmetric, $J_{ij}=J_{ji}$, then this dynamics obeys detailed balance because it represents noisy gradient descent $\partial_t \phi_i =
- \partial_{\phi_i}  H + \zeta_i(t)$ on the energy function
\be
H=\sum_i \left(  \frac{\mu}{2} \phi^2_i +\frac{g}{4!}  \phi^4_i \right) - \frac{1}{2}\sum_{ij} J_{ij} \phi_i \phi_j -\sum_i h_i 
\label{eq:H}
\phi_i
\ee
which can be thought of as a soft spin version of the Sherrington-Kirkpatrick model \cite{sherrington1975solvable}. 
The diagrammatic technique in its MSR incarnation can be applied irrespective of any such restriction, i.e.\ whether or not the system obeys detailed balance. For a supersymmetric treatment interaction symmetry is necessary, on the other hand.

The  generating functional has the form\eq{Z_av} with action, written directly in the continuous time limit and using the Ito convention,
\bea
%&&Z=\int D\phi D\hat{\phi} \exp[-S]\\
S&=&\int dt \sum_i \left [ T \hat{\phi}_i^2 + i  \hat{\phi}_i \left ( \partial_t \phi_i +\mu \phi_i + \frac{g}{3!}  \phi^3_i -\sum_j J_{ij} \phi_j\right)\right]
\label{eq:action_SK}
\eea
In this action the interaction gives an additional vertex with two legs, $\int dt\,\sum_{ij} J_{ij}\hat\phi_i(t)\phi_j(t)$. (As before we do not introduce a new symbol for this vertex as the meaning is clear from the number of legs attached.) The diagrammatic expansion now becomes a joint expansion in both $g$ and the interaction amplitude; formally one could set $J_{ij}=J\hat{J}_{ij}$, consider the $\hat{J}_{ij}$ fixed and expand in $J$. To illustrate the new diagrams that appear we restrict ourselves to $g=0$ (and $h=0$). The expansion of the propagator to second order in $J$ is then simply
\begin{eqnarray*}
\parbox{12mm}{\begin{fmfgraph*}(10,20)
\fmfleft{i}
\fmfright{j}
\fmf{dbl_plain}{i,j}
\end{fmfgraph*}}
&=& 
\
\parbox{12mm}{\begin{fmfgraph*}(10,20)
\fmfleft{i}
\fmfright{j}
\fmf{plain}{i,j}
\end{fmfgraph*}}
+\ 
\parbox{12mm}{\begin{fmfgraph*}(10,20)
\fmfleft{i}
\fmfright{j}
\fmfdot{n}
\fmf{plain,tension=0.5}{i,n,j}
\end{fmfgraph*}}
+\ 
\parbox{15mm}{\begin{fmfgraph*}(15,20)
\fmfleft{i}
\fmfright{j}
\fmfdot{m,n}
\fmf{plain,tension=0.5}{i,m,n,j}
\end{fmfgraph*}}
\ + \order(J^3)
\end{eqnarray*}
All propagators and vertices now carry site indices in addition to the time index and the ``sector'' label 1 or 2 for physical ($\phi$) and conjugate ($\hat\phi$) fields. If we focus on the response function part of the propagator, the expansion becomes
\begin{eqnarray}
\parbox{15mm}{\begin{fmfgraph*}(13,20)
\fmfleft{i}
\fmfright{j}
\fmf{dbl_plain_arrow}{i,j}
\end{fmfgraph*}}
&=& 
\
\parbox{12mm}{\begin{fmfgraph*}(10,20)
\fmfleft{i}
\fmfright{j}
\fmf{plain_arrow}{i,j}
\end{fmfgraph*}}
+\ 
\parbox{12mm}{\begin{fmfgraph*}(10,20)
\fmfleft{i}
\fmfright{j}
\fmfdot{n}
\fmf{plain_arrow,tension=0.5}{i,n,j}
\end{fmfgraph*}}
+\ 
\parbox{15mm}{\begin{fmfgraph*}(15,20)
\fmfleft{i}
\fmfright{j}
\fmfdot{m,n}
\fmf{plain_arrow,tension=0.5}{i,m,n,j}
\end{fmfgraph*}}
\ + \order(J^3)
\end{eqnarray}
Writing out the diagrams this translates to
\begin{eqnarray}
R_{ij}(t,t') &=& R_{ij0}(t,t')
+ \int dt_1R_{il0}(t,t_1) J_{lk} R_{kj0}(t_1,t') \\
&&{}+ \int dt_1 dt_2 R_{in0}(t,t_2) J_{nm} R_{ml0}(t_2,t_1)J_{lk} R_{kj0}(t_1,t') + \order(J^3)
\nonumber
\end{eqnarray}
where all internal site indices ($k,l,m,n$) are to be summed over. Here $R_{ij0}(t,t')$ is the response function of the unperturbed dynamics, which because of the absence of interactions is diagonal in the site indices, $R_{ij0}(t,t')=\delta_{ij}R_0(t,t')$. 
The time integrals become simple products in frequency space, giving for the Fourier transform $R_{ij}(\omega) = \int dt\, R_{ij}(t,t')e^{i\omega (t-t')}$
\begin{eqnarray}
R_{ij}(\omega) &=& R_{ij0}(\omega)
+ R_{il0}(\omega) J_{lk} R_{kj0}(\omega) 
\label{response_ij}\\
&&{}+ R_{in0}(\omega) J_{nm} R_{ml0}(\omega)J_{lk} R_{kj0}(\omega) + \order(J^3)
\nonumber
\end{eqnarray}
or in matrix form
\begin{eqnarray}
\bm{R}(\omega) &=& \bm{R}_0(\omega) + \bm{R}_0 (\omega) \bm{J} \bm{R}_0(\omega) + 
\bm{R}_0(\omega) \bm{J} \bm{R}_0(\omega)
\bm{J} \bm{R}_0(\omega) + \order(J^3)\\
&=& 
[\bm{R}_0^{-1}(\omega) - \bm{J}]^{-1}
\end{eqnarray}
Inserting $\bm{R}_0^{-1}(\omega) = (\mu-i\omega)\mident$ where $\mident$ is the identity matrix gives $\bm{R}(\omega)=[(\mu-i\omega)\mident-\bm{J}]^{-1}$. 
Given that we are considering $g=0$ where the dynamics is purely linear, this is easily seen to be the exact result. This was possible to obtain here because we were able to sum up all diagrams, or equivalently because only a single diagram (the bare quadratic vertex) contributes to the self-energy.

Up to here the discussion applies for general $J_{ij}$. To simplify further one needs to make assumptions on the statistics of these interactions. 
One interesting case is that of a soft-spin Sherrington-Kirkpatrick model for which the $J_{ij}$ are 
zero-mean Gaussian random variables, uncorrelated for different index pairs $ij$, except that the symmetry in the interaction matrix is imperfect:
\be 
\langle J_{ij}^2\rangle = \frac{J^2}{N}, \qquad
\langle J_{ij}J_{ji}\rangle = \frac{\kappa J^2}{N}
\label{eq:Jij_statistics}
\ee
The symmetry parameter has the value $\kappa=1$ for a fully symmetric matrix, while $\kappa=0$ gives a fully asymmetric matrix.

For the local response function $R_{ii}$ one can in general simplify\eq{response_ij} to
\begin{eqnarray}
R_{ii}(\omega) &=& R_{0}(\omega)
+ R_{0}(\omega) J_{im} R_{0}(\omega)J_{mi} R_{0}(\omega) + \order(J^3)
\end{eqnarray}
where the term of first order in $J_{ij}$ vanishes due to the lack of self-interactions. 
For the soft-spin SK model, the sum $\sum_m J_{im}J_{mi}$ in the second order term has average $(N-1)\kappa J^2/N = \kappa J^2 + \order(1/N)$ while its variance is $\order(1/N)$. For large $N$ it is therefore self-averaging, i.e.\ equal to $\kappa J^2$ with probability one. Hence
\begin{eqnarray}
R_{ii}(\omega) &=& R_{0}(\omega)
+ \kappa J^2 R_{0}^3(\omega) + \order(J^3)
\label{Rii_OJ2}
\end{eqnarray}
As one might have expected, because this model has each node interacting with all others the nodes become equivalent, making the local response functions $R_{ii}$ independent of $i$. It remains true at higher orders that the local response only depends on the overall coupling amplitude $J$ and the symmetry parameter $\kappa$. This can be shown by a separate diagrammatic argument \cite{feigel1979phase,hertz1979dynamics,verbaarschot1984evaluation} or by explicit averaging of the path generating function $Z$ over the disorder, i.e.\ the statistics of the $J_{ij}$. This latter approach is the subject of the following section.

\section{Quenched averaged dynamics of soft spin models}
\label{sec:quenched_average}

In the previous sections, we described how the dynamics of a set of interacting scalar variables $\phi_i$ that evolve according to a stochastic differential equation
can be studied perturbatively using the path integral formalism. When the evolution equations depend on random quantities such as the $J_{ij}$ above, then instead of studying a single system with fixed $J_{ij}$ one can consider an ensemble of systems, each with a different set of interactions $J_{ij}$ drawn from some distribution. The resulting ensemble averages are expected to reflect the behavior of a typical sample as far as macroscopic quantities such as the average local response $(1/N)\sum_i R_{ii}$ are concerned: these are {\em self-averaging}, i.e.\ to leading order dependent only on the overall statistics of the interactions in the system. Where the interactions are weak and long-ranged as in the soft-spin SK model, all spins also become equivalent so even the local responses $R_{ii}$ are self-averaging as we saw above. This would not be true, for example, for systems with interactions on a network with finite connectivity.

We illustrate the approach for the dynamics given by~(\ref{eq:phiint}), with dynamic action in \eref{eq:action_SK}
and assuming Gaussian statistics for the $J_{ij}$ as specified in \eref{eq:Jij_statistics}. %\red{
The derivation that we outline here was first described in \cite{sompolinsky1981dynamic,sompolinsky1982relaxational} for symmetric interactions; the role of the degree of asymmetry in the interactions was analysed later in \cite{crisanti1987dynamics}.

Using the fact that the average of $\exp(i\bm{x}\cdot\bm{z})$ over a vector of zero mean Gaussian random variables $\bm{z}$ with covariance $\bm{A}^{-1}$ is $\exp(-\bm{x}\cdot\bm{A}^{-1}\bm{x}/2)$,
the $\bm{J}$-average (denoted by an overline) of the part of the generating functional $Z$ that depends on the $J_{ij}$ reads
\bea
\fl I&\equiv& \prod_{i<j} \overline{\exp\left [ i  \int dt \, (\hat{\phi}_i \phi_j J_{ij} +\hat{\phi}_j \phi_i J_{ji})\right]}
\\
\fl &=&\prod_{i<j} 
\exp\left\{-\frac{J^2}{2N}\int dt\,dt'\,\left[\hat\phi_i(t)\hat\phi_i(t')\phi_j(t)\phi_j(t')
+  \hat\phi_j(t)\hat\phi_j(t')\phi_i(t)\phi_i(t')\right.\right.
\nonumber\\
\fl &&
\left.\left.{}
+2\kappa \hat\phi_i(t)\phi_j(t)\hat\phi_j(t')\phi_i(t')
\right]
\right\}
\\
\fl &=&\prod_{i\neq j} 
\exp\left\{-\frac{J^2}{2N}\int dt\,dt'\,\left[\hat\phi_i(t)\hat\phi_i(t')\phi_j(t)\phi_j(t')
+\kappa \hat\phi_i(t)\phi_i(t')\phi_j(t)\hat\phi_j(t')
\right]
\right\}
\eea
One can drop the restriction $i\neq j$ for large $N$ as including the $i=j$ terms only gives subleading corrections.  A mean-field decoupling of the quartic terms gives
\bea
I &=& \prod_i \exp(-L_i)
\\
L_i &=& 
\frac{J^2}{2} \int dt\,dt' [C(t,t')\hat{\phi}_i(t)\hat{\phi}_i(t') - i \kappa R(t,t') \hat{\phi}_i(t)\phi_i(t')]
\eea
Here
\bea
C(t,t')&\equiv& \frac{1}{N} \sum_{j} \phi_j(t) \phi_j(t'),\ \ \ \ R(t,t')\equiv  \frac{i}{N} \sum_{j} \phi_j(t) \hat{\phi}_j(t') 
\eea
are the average local correlation and response functions of the system. They are in principle dependent on the specific trajectory of the system, but from the law of large numbers will assume deterministic values for $N\to\infty$. This heuristic argument can be justified by a formal calculation introducing conjugate order parameters to $C$ and $R$ and making a saddle point argument\cite{sompolinsky1981dynamic,sompolinsky1982relaxational,crisanti1987dynamics}. Because $I$ now factorizes over sites $i$, so does the entire ensemble-averaged partition function $\overline{Z}$. The contributions from $L_i$ give an effective noise with correlation function $J^2 C(t,t')$, and a delay term with memory kernel $\kappa R(t,t')$.  

All spins have the same action, so we can drop the index $i$ and write down the effective dynamics as
\newcommand{\zetaeff}{\zeta_{\rm eff}}
\bea
\partial_t \phi= - \mu \phi - \frac{g}{3!}  \phi^3+\kappa J^2 \int^t_{0} dt' R(t,t') \phi(t') +h(t) +  \zetaeff(t)\\
\langle \zetaeff(t) \zetaeff(t')\rangle = 2T \delta(t-t')+J^2 C(t,t')
\eea
The correlation and response need to be found self-consistently from this, and to facilitate calculation of the response we have added back in the field term $h(t)$.

Let us now take a moment to study the dynamics of the bare model ($g=0$) that arises from the above equation. At $g=0$, and assuming that for long times a stationary regime is reached where $R(t,t')=R(t-t')$, the equation of motion in frequency space is
\be
-i \omega \phi(\omega)= - \mu \phi(\omega) + \kappa J^2 R(\omega)\phi(\omega) + h(\omega)+\zetaeff(\omega)
\ee
Averaging both sides over the noise term $\zetaeff$ gives for the mean $m(\omega)=\langle \phi(\omega)\rangle$
\be
(-i\omega + \mu - \kappa J^2 R(\omega)m(\omega) =
h(\omega)
\ee
From this, we find the response function
\be
\fl R^{-1}(\omega)= \left[\frac{\partial  m(\omega) }{\partial h(\omega)}\right]^{-1}=
\frac{\partial  h(\omega) }{\partial m(\omega)}
=
-i\omega+\mu-\kappa J^2 R(\omega)
= R_0^{-1}(\omega) - \kappa J^2 R(\omega)
\label{eq:R0}
\ee
To leading order in $J^2$ this gives
$
R^{-1}(\omega)= R_0^{-1}(\omega) -\kappa J^2 R_0(\omega)
$,
which after inverting and re-expanding to $\order(J^2)$ agrees with the perturbative result\eq{Rii_OJ2} as it should.

The critical dynamics of system can be understood via the low frequency behavior of the response function. In fact, one can define a characteristic response time scale as 
\begin{equation}
\tau = \frac{\int dt \,(t-t') R(t-t')}{\int dt\, R(t-t')}
= \left. \frac{1}{iR} \frac{\partial R}{\partial \omega}\right|_{\omega=0}
\end{equation}
Taking the $\omega$-derivative of both sides of (\ref{eq:R0}) and multiplying by $R$ gives
\bea
-\left(\frac{1}{R}\frac{\partial R}{\partial \omega}\right) = -iR - J^2 R^2 
\left(\frac{1}{R}\frac{\partial R}{\partial \omega}\right)
\eea
and evaluating at $\omega=0$ one finds
\bea
&&-i\tau = -iR(0)-iJ^2 R^2(0)\tau
\eea
The response timescale can therefore be expressed as
\be
\tau = \frac{R(0)}{1-J^2 R^2(0)}
\ee 
In other words, when $R(0)=J^{-1}$, the relaxation time scale of the system diverges and the system exhibits critical slowing down. From (\ref{eq:R0}) one sees that the critical value of $J$ obeys $J = \mu - J^2/J$, hence $J=\mu/2$. 
Using the explicit solution of (\ref{eq:R0}), which reads
\be 
R(\omega) = (2J^2)^{-1}\left[\mu-i\omega - \sqrt{(\mu-i\omega)^2-4J^2}\right]
\ee
shows further that at criticality the response function has a singularity $\propto \omega^{1/2}$ for $\omega\to 0$. In the time domain this corresponds to a power law tail $R(t-t') \propto (t-t')^{-3/2}$, which is responsible for the diverging mean timescale.

\iffalse
The critical dynamics of system can be understood via the low frequency behavior of the response function. In fact, we can define a damping coefficient $\Gamma$ as
\begin{equation}
\Gamma^{-1} \equiv i \frac{\partial R^{-1}}{\partial \omega}
\end{equation}
that describes is the time scale of the response of the system. Taking the derivatives of both sides of Eq.\ \ref{eq:R0} and evaluating it at $\omega=0$ we can find that
\bea
&&\Gamma_0=\lim_{\omega \to 0} \left (\frac{\partial R^{-1}_0}{\partial (\omega )}\right) 
= 1+ J^2 R^2_0(0) \left[ \lim_{\omega \to 0} \left (\frac{\partial R^{-1}_0}{\partial (\omega )}\right) \right ]\nonumber\\
&&=1+J^2 R^2_0(0) \Gamma^{-1}_0 \
\eea

Consequently, the response time scale would be
\be
\tau = \Gamma^{-1}_0=\frac{1}{1-J^2 R^2_0(0)}
\ee 

In other words, at $R(0)=J^{-1}$ the relaxation time scale of the system diverges and we face a critical slow-down.
\fi

\section{Path integrals for hard spin models}

So far we have considered models in which the dynamical variables took continuous values. Here we show that, with some modifications, the same approach can be used to study the dynamics of models involving Ising spins.

Consider a system of $N$ binary spin variables, $\bsig=(\sigma_1,\ldots, \sigma_N)$ with $\sigma_i=\pm 1$. We consider synchronous dynamics, where time $t$ advances in discrete steps and at each time step all spins are updated. Specifically, we assume that the spins decide their states according to the following probability distribution
\bea
&&P[\bsig(t+1)|\bsig(t)] = \prod_i \frac{\exp[h_i(t)\sigma_i(t+1)]}{2 \cosh(h_i(t))} \label{synch-def}
\\
&&h_i(t) = h^{\rm ext}_i+\sum_j J_{ij}\sigma_j (t).\label{field-def}
\eea
If the external fields $h^{\rm ext}_i$ are constant in time as written above and if the couplings are symmetric, i.e.\  $J_{ij}= J_{ji}$, this dynamics reaches an equilibrium state where detailed balance is satisfied and  configurations are visited according to the Boltzmann distribution $p(\bm \sigma) \propto \exp(-H)$ with the Hamiltonian \cite{peretto1984collective}
\be
H=- \sum_i h^{\rm ext}_i \sigma_i - \sum_i \lc(h_i(\bm \sigma)),
\ee
where we have introduced the abbreviation $\lc(x)  \equiv \ln(2 \cosh(x))$ and emphasize that it is the total fields $h_i(\bm \sigma)$, which depend on the configuration, that appear inside the log $\cosh$.

An alternative to synchronous dynamics that is commonly studied in the literature is
continuous time Glauber dynamics \cite{glauber1963time}, in which spins are updated individually and at random (exponentially distributed) time intervals. This is achieved by prescribing that in any infinitesimal time interval $\delta t \to 0$, every spin is flipped with probability
\be
\delta t \left[1-\frac{1}{2} \sigma_i(t) \tanh(h_i(t))\right].
\ee
Again, for constant external fields and symmetric couplings an equilibrium Boltzmann distribution is reached, but this time with the familiar Ising Hamiltonian
\be
H = - \sum_i h^{\rm ext}_i \sigma_i - \sum_{i<j} J_{ij} \sigma_i \sigma_j.
\ee
For derivations of these equilibrium distributions see \cite{coolen2001statistical,coolen_book}.
Although the Glauber dynamics is the one that leads to the usual Ising Hamiltonian, in what follows we focus on the synchronous case for two reasons: (i) the path integral formulation is slightly simpler in terms of notation and (ii) the synchronous case has been the focus of recent work on the dynamics of hard spin models, including our own on the inverse problem of inferring the couplings in the model from the statistics of the spin history \cite{roudi2011mean,mezard2011exact,mahmoudi2014generalized}. 

\subsection{Path integral formulation}

The generating functional for the dynamics is defined as
\bea
Z[\psi] =\left\langle \exp\bigg (\sum_i\psi_i(t)\sigma_i(t)\bigg)\right\rangle,
\eea
where the average denoted by $\langle \cdots \rangle $  is over the distribution of trajectories generated according to the probability distribution\eq{synch-def} and the $\psi_i$ are fields that allow one to obtain the statistics of the $\sigma_i$ by taking derivatives. The key to writing a path integral representation of the generating functional in this case is to work with the local fields $h_i(t)$, which are continuous for $N\to\infty$, and not the original spin variables:
\be
\fl Z[\psi]=\Tr \int D[h] \prod_{t} \delta(\bm h(t)-\bm h^{\rm ext}-\bm J \bm \sigma (t)) \exp\bigg( \sum_t[\bm \psi(t+1)+\bm  h(t)]\cdot \bm  \sigma(t+1)-\sum_{i,t} \lc(h_i(t))\bigg) \label{eq:Zdelta}
\ee
where $\bm J$ is the interaction matrix, $\Tr$ indicates a sum over all spin trajectories $\{\bm \sigma(t)\}$ and similarly $D[h]$ an integral over all field trajectories $\{\bm h(t)\}$. The discrete time variable range is $t=0,1,\ldots,T-1$ where $T$ is the final time.

The delta function in \eref{eq:Zdelta} is introduced to enforce the definition \eref{field-def} and can be written as a Fourier transform, leading to
\bea
\fl Z[\psi]&=&\Tr \int D[h\hat h] \exp(-S_\sigma)\label{eq:trZ}\\
\fl S_{\sigma}&=&-\sum_t i \hat{\bm h}(t)\cdot[\bm h(t)-\bm h^{\rm ext}-\bm J \bm \sigma (t)]-\sum_t[\bm \psi(t+1)+\bm h(t)]\cdot \bm \sigma(t+1)+\sum_{i,t} \lc(h_i(t)).
\eea
This expression can be used in several ways. One is to average over the distribution of $\bm{J}$; the second is to derive mean-field equations for the system. We do not go into the first route here, i.e.\ the quenched averaged dynamics of the system, as this is both very involved and also discussed in detail elsewhere~\cite{coolen2000statistical}. However, as an example of how one can use the path integeral formulation \eref{eq:trZ} of the generating functional, we consider in the following subsections the simple saddle point approximation to the path integral as well as possible improvements to this.

\subsection{Saddle point} 

To derive the saddle point equations, we first perform the trace in \eref{eq:trZ}, which for a single timestep involves
\bea
\fl &&
\!\!\!\!\!\!\!\!\!\!\!\!\!\!\!
\sum_{\bm\sigma(t)} \exp[-i\hat{\bm h}(t)\cdot \bm J\bm \sigma (t)+(\bm \psi(t)+\bm h(t-1))\cdot \bm \sigma(t)] 
\nonumber\\
&=&
\prod_i \sum_{\sigma_i=\pm 1} \exp\bigg(\Big[\psi_i(t)+h_i(t-1)-i\sum_j \hat{h}_j(t)J_{ji}\Big]\sigma_i(t)\bigg)\\
&=&\prod_i 2\cosh(\psi_i(t)+h_i(t-1)-i (\bm J^{\rm T}\hat{\bm h}(t))_i).
\eea
Using this we can write
\bea
\fl Z[\psi]&=& \int D[h\hat h] \exp(-S)\\
\fl S&=& -\sum_t i \hat{\bm h}(t)\cdot[\bm h(t)-\bm h^{\rm ext}]-\sum_{i,t} \left[
\lc\big(\psi_i(t+1)+h_i(t)-i(\bm J^{\rm T}\hat{\bm h}(t+1))_i\big) -\lc(h_i(t))\right].
\label{eq:action-after-trace}
%\nonumber 
\eea
The saddle point equations for stationarity of $S$ with respect to the $h_i(t)$ and $\hat h_i(t)$ are then
\bea
%\fl 
\fl \frac{ \partial S}{\partial h_i(t)}= -i\hat{h}_i(t)-\tanh\big(\psi_i(t+1)+h_i(t)-i(\bm J^{\rm T} \hat{\bm h}(t+1))_i\big)+ \tanh(h_i(t))=0
\label{eq:h-stationarity}\\
\fl \frac{\partial S}{\partial \hat{h}_i(t)}=- i[h_i(t)-h_i^{\rm ext}]+i \sum_j J_{ij} \tanh\big (\psi_j(t)+h_j(t-1)-i(\bm J^{\rm T}\hat{\bm h}(t))_j\big )=0.
\eea
These equations can be written in a simpler form in terms of the 
magnetizations in the system biased by $\psi$, which are generally given by $m_i(t)=\partial_{\psi_i(t)} \ln Z$. In the saddle point approximation $Z\approx Z_s\equiv \exp(-S)$, and because $S$ is stationary with respect to $h$ and $\hat h$,
\be
m_i^s(t)=-\partial_{\psi_i(t)} S
= \tanh\big (\psi_i(t)+h_i^s(t-1)-i(\bm J^{\rm T}\cdot \hat{\bm h}^s(t))_i\big) \equiv \mu_i(t)
\label{eq:mu_def}
\ee
Here we have used ``s'' superscripts to denote saddle point values and introduced $\mu_i(t)$ as a convenient abbreviation for later use. The saddle point equations then simplify to
\bea
i \hat{h}_i^s(t)+m_i^s(t+1)- \tanh(h_i^s(t))=0
\label{eq:sp1}\\
h_i^s(t)=h_i^{\rm ext}+\sum_j J_{ij} m_j^s(t).
\label{eq:sp2}
\eea
For the physical dynamics we want the solution at $\psi=0$, which we denote with a ``0'' superscript.
One can show that this has $\hat h_i^0(t)=0$. Indeed, if one considers the dynamics over a finite number of timesteps $t=1,2,\ldots,T$, then both $\bm h(t)$ and $\hat{\bm h}(t)$ are defined over the range $t=0,1,\ldots,T-1$ and accordingly one finds that in the first saddle point equation\eq{eq:h-stationarity} taken at $t=T-1$, the $\hat{\bm h}(t+1)$-term inside the tanh is absent. For $\psi=0$ this equation then dictates $\hat h_i^0(T-1)=0$, and working backwards in time from there one finds recursively $\hat h_i^0(t)=0$ for all $t$. Accordingly \eref{eq:sp1} and \eref{eq:sp2} simplify to the standard mean field equations for the magnetizations,
%From the definition of the saddle point solution for the mean magnetizations, $m^s_i$, is derived from the saddle point equations at $\psi=0$, therefore admit the following solutions 
\bea
&&m^0_i(t+1)= \tanh(h_i^0(t))
\label{eq:naive_mf1}
\\
%&&\hat{h}^s_i(t)=0\\
&&h^0_i(t)=h^{\rm ext}_i+\sum_j J_{ij} m^0_j(t).
\label{eq:naive_mf2}
\eea
Note that the saddle point value of the action \eref{eq:action-after-trace} is then zero for $\psi = 0$, as expected from the normalization condition $Z[0]=1$. 

\subsection{Beyond the saddle point: naive approach}
One can go beyond the saddle point approximation and take into account the Gaussian fluctuations around the saddle point to obtain a better estimate of the generating functional and hence of the equations of motion for the $m_i(t)$. The Gaussian corrections to the log generating functional are
\be
A[\psi]= \ln \int D[h\hat{h}] \exp\left(- \frac{1}{2} [\delta h\ \hat{h}]\T \cdot  \partial^2 S [\delta h\ \hat{h}]\right) = -\frac{1}{2} \ln |\partial^2 S|,
\ee
where $\delta h$ indicates the deviation of $h$ from its saddle point value.  
We denote by $\partial^2 S$ the matrix of the second derivatives of $S$ with respect to $h$ and $\hat{h}$ calculated at the saddle point; this has entries 
\begin{eqnarray}
\frac{\partial^2 S}{\partial h_i(t) \partial h_j(t')}=\delta_{ij} \delta_{tt'} \left[\mu^2_{i}(t+1)-\tanh^2(h_i^s(t))\right]
\label{eq:hessian_h}
\\
\frac{\partial^2 S}{\partial h_i(t) \partial \hat{h}_j(t')}=-i\delta_{ij}\delta_{tt'}+i \delta_{t+1,t'} J_{ji} [1-\mu^2_i(t+1)]
\label{eq:hessian_cross_terms}\\
\frac{\partial^2 S}{\partial \hat{h}_i(t) \partial \hat{h}_j(t')} =\delta_{tt'} \sum_k J_{ik} J_{jk} [1-\mu^2_k(t)],
\label{eq:hessian_hhat}
\end{eqnarray}

From the corrected generating functional $Z[\psi]=Z_s[\psi]+A[\psi]$ we obtain a corrected expression for the magnetization 
\be
m_i(t) = \left.
\frac{\partial \ln Z[\psi]}{\partial \psi_i(t)}
\right|_{\psi=0}
 = m^0_i(t)+\left.
 \frac{\partial A[\psi]}{\partial \psi_i(t)}
 \right|_{\psi=0}  \label{eq:mcorrected}.
\ee
To calculate the $\psi$-derivative of $A$ at $\psi=0$ we can either appeal to numerical methods or evaluate $A$ approximately. One relatively simple approximation, which will give us an intuitive feeling for the corrections and turns out to capture the lowest order corrections in $J$, is to only keep the equal-time ($t=t')$ elements of the matrix $\partial^2 S$, i.e.\ to discard the second term in\eq{eq:hessian_cross_terms}. The matrix then separates into blocks corresponding to the different timesteps $t$. Each of those blocks is of size $2N\times 2N$. Let us order the elements so that the top left $N\times N$ sub-block contains the entries from\eq{eq:hessian_h}. This block, which we will denote by $\bm{\alpha}_t$, is diagonal and vanishes at the physical saddle point, where $\psi_i(t)=0$ and $\hat{h}_i^s(t)=\hat{h}_i^0(t)=0$. The two off-diagonal blocks\eq{eq:hessian_cross_terms} are $-i\mident$ within our approximation. If we call the bottom right $N\times N$ block $\bm{\beta}_t$, then we have by a standard determinant block identity
\be
A[\psi]= -\frac{1}{2} \sum_t \ln |\mident + \bm{\alpha}_t \bm{\beta}_t|
=
-\frac{1}{2} \sum_t {\rm Tr} \ln(\mident + \bm{\alpha}_t \bm{\beta}_t)
\label{eq:A_block_diagonal}
\ee
Near the physical saddle point $\bm{\alpha}_t$ is small so we can linearize the logarithm to get
\bea
A[\psi]&\approx & -\frac{1}{2} \sum_t {\rm Tr}(\bm{\alpha}_t \bm{\beta}_t)
\\
&=& -\frac{1}{2}\sum_{it} [\mu^2_{i}(t+1)-\tanh^2(h^s_i(t))] \sum_k J^2_{ik} [1-\mu^2_k(t)].
\label{eq:A_linearized}
\eea
Linearizing then also the diagonal entries of ${\bm \alpha}_t$ that appear in the first square bracket, and accordingly setting the last factor to its value at the physical saddle point gives
\bea
\fl A[\psi]
&\approx & - \sum_{it} \big (\psi_i(t+1)-i(\bm J^{\rm T}\cdot \hat{\bm h}^s(t+1))_i\big) 
[1-m_i^0(t+1)^2] m_i^0(t+1) 
\sum_k J^2_{ik} [1-m_k^0(t)^2].
\eea
When we take the derivative of this with respect to $\psi_i(t)$, we do in principle get a term from the dependence of $\hat{\bm h}^s$ on $\psi$. But as $\hat{\bm h}^s$ is multiplied by a factor of $\bm J$, this will give us a contribution to the derivative that is of higher order in $J$ than the main term from the explicit dependence on $\psi_i(t)$. Discarding this higher order contribution -- as our second approximation in addition to neglecting correlations between different time steps in $\partial^2 S$ -- yields for the derivative
\bea
A_i(t) \equiv \frac{\partial A[\psi]}{\partial \psi_i(t)} &\approx&
- [1-m_i^0(t)^2] m_i^0(t) \sum_k J^2_{ik} [1-m_k^0(t-1)^2]
\label{eq:Ai}
\eea
Here we have implicitly already set $\psi= 0$ because we linearized around the physical saddle point throughout.
Using this result in \eref{eq:mcorrected}, we obtain the corrected magnetization as
\be
m_i(t) = m^0_i(t) - \left [1-m^0_i(t)^2 \right] \left \{ m^0_i(t) \sum_k J^2_{ik} [1-m^0_k(t-1)^2 ] \right \}.
\label{eq:laplacesimple}
\ee
This is the result of our naive approach to including Gaussian corrections in the saddle point integral. Note that because of the normalization $Z[0]=1$, we do not expect Gaussian corrections to $\ln Z[0]$ at the physical saddle point. The approximation we made in evaluating $A[\psi]$ as a sum of local-in-time terms preserves this requirement: $A[0]=0$ as is clear from\eq{eq:A_block_diagonal} given that $\bm{\alpha}_t=0$ when $\psi=0$.

\subsection{Beyond the saddle point: Plefka and Legendre transform approaches}

It is instructive to compare\eq{eq:laplacesimple} with the so-called dynamical TAP equations for our system \cite{roudi2011dynamical}
\be
\fl m^{\rm TAP}_i(t) = \tanh\left( 
h^{\rm ext}_i+\sum_j J_{ij} m_j^{\rm TAP}(t)
%h^s_i(t-1)
-  m^{\rm TAP}_i(t) \sum_k J^2_{ik} [1-m^{\rm TAP}_k(t-1)^2 ] \right),
\label{eq:DynTAP}
\ee
which for stationary magnetizations reduce to the better-known equilibrium TAP equations \cite{kappen2000mean,thouless1977solution}. The negative term inside the hyperbolic tangent is known as the Onsager correction, which improves on the naive estimate of the mean effective field acting on a spin.

One now observes that the term inside the curly brackets on the right hand side of our simpler corrected equation\eq{eq:laplacesimple} is exactly of the form of this Onsager correction, so there is a close relation between the two approaches.
However, there are two significant differences. 
First, while the Onsager term in \eref{eq:DynTAP} appears inside the $\tanh$, in \eref{eq:laplacesimple} it is outside, as if we had linearized in the correction. This is important:
it means there is nothing on the right hand side of \eref{eq:laplacesimple} to stop the magnetizations from going outside the physical range  $[-1,1]$. The origin of this difference is the fact that while the dynamical TAP method corrects the naive estimate of the {\em field} acting on a spin, the Gaussian fluctuation approach described in the previous subsection directly estimates corrections for the magnetizations themselves.

The second difference between \eref{eq:laplacesimple} and \eref{eq:DynTAP} is that in the latter
it is the corrected magnetizations $m^{\rm TAP}_i$ itself that appear in the Onsager term, making the approximation self-consistent. In our naive method, on the other hand, the Onsager term is evaluated at the uncorrected saddle-point magnetizations.

The dynamical TAP equations \eref{eq:DynTAP} can be derived by the Plefka approach \cite{plefka1982convergence}. This is an elegant method that captures corrections to the effective fields as explained above, ensuring in contrast to \eref{eq:laplacesimple} that the magnetizations remain in the physical range even when corrections to the naive mean field estimate are large. 
This is achieved by working not with the generating functional directly (or the Helmholtz free energy in the equilibrium case), but with its Legendre transform with respect to the magnetizations (the equivalent of the equilibrium Gibbs free energy).
In this way, one essentially approximates the system with an independent spin model in which each spin feels an effective field $h^{\rm eff}_i$. This leads to expressions of the form $m_i(t) = \tanh(h^{\rm eff}_i(t-1))$ for the magnetizations, which therefore always lie between $-1$ and $1$. The effective fields are determined so as to best match the Gibbs free energy of the original model, by a perturbation expansion to second order in $J$.

Interestingly, the philosophy of working with the Legendre transform can also be applied to the approach we have described above, which initially has a rather different starting point, namely a saddle point approximation with Gaussian corrections. We show this in the Appendix, where we demonstrate that by switching to the Gibbs free energy and keeping only first order terms in the corrections $A_i$ one can retrieve exactly the dynamical TAP equations. We refer the reader to~\cite{negele1988quantum} and~\cite{kholodenko1990onsager} for applications of the same idea in the context of the equilibrium Ising and Potts models. 

\section{Summary and Discussion}

Path integral methods and the associated perturbative and diagrammatic tools are among the most powerful methods used in almost every part of theoretical physics. Disordered systems, such as spin glasses, neural networks, models of the immune system etc.\ are no exception and, over the years, path integrals have played a significant role in the study of equilibrium and out-of-equilibrium properties of these systems. 
In disordered systems, by definition, interactions in a single system do not have any trivial structure; there is symmetry only in their statistics, as described by the distribution from which they are drawn.  These systems are typically subject to external stochastic forces with both rapidly changing and quenched random components. Furthermore, these sources of randomness are expected to play crucial roles in the physics and should therefore be included explicitly in studying such systems.  The path integral and diagrammatic methods treated in textbooks on field theoretical tools for other physical systems usually do not deal with disordered systems.  Therefore, we tried in this paper to describe some of the key field-theoretic and path integral techniques in a manner that can be applied directly to disordered systems.

We started by studying the dynamics of a scalar degree of freedom evolving according to a Langevin equation and showed how one can do perturbative calculations for this model and represent them in diagrammatic form. We then discussed the supersymmetries of the action that features in the path integral description of this system, the physical intuition and meaning behind them, and how they can help in doing diagrammatic perturbation theory by reducing the graphs to be considered to those in simpler, well-understood equilibrium calculations. Our next step was to study the dynamics of systems of interacting ``soft-spin'' variables subject to Langevin equations with random interactions, first using a perturbative treatment for a single sample of the interactions, and then via a conceptually different approach based on averaging the generating function in its path integral formulation over the disorder.

Finally we switched to hard-spin models, focusing on the Ising model with synchronous update dynamics, for a single sample with arbitrary couplings. Here, as opposed to the way we treated soft-spin models, the path integral is not written in terms of the spins directly but in terms of the fields acting on them. We discussed how, from such a path integral formulation, one can derive approximate equations for the mean magnetizations at the saddle point level  and compute the Gaussian fluctuation corrections around the saddle point. We showed that a naive calculation of these corrections yields equations of motion for the magnetizations that can lead to unphysical predictions. These issues can be avoided by going to improved approximation schemes like dynamical TAP equations, as  described recently in other papers using a path integral formulation \cite{roudi2011dynamical,Bachschmid-Romano2016extended}, as well as several alternative approaches \cite{kappen2000mean,aurell2012dynamic,mahmoudi2014generalized}. We closed the paper with the intriguing observation, however, that issues with naively corrected mean-field equations can also be cured within the general approach presented in this paper, using a Legendre transform to switch to the Gibbs free energy.
%corrs In the last section, however, we illustrated a general strategy for avoiding the unphysical properties by pursuing a slightly different approach based on Legendre-transforming the logarithm of the generating function, following ideas that were used in the equilibrium case in \cite{negele1988quantum} and \cite{kholodenko1990onsager}. 
It is tempting to infer that the Legendre transform implicitly achieves a resummation of the most important diagrams, but we have not been able to show this explicitly and leave this point as an open question for future work.

Although we have tried to cover what we view as the key concepts in this review, we have had to leave out several important issues to maintain a coherent focus. We list two of these in the following.

%a number of important issues outside this review, as they would not fit in the main scope of the review and would have significantly lengthen the paper, but we find it useful to briefly mention them here.

{\em The use of dynamical models in inference.}
The path integral approach described here is designed initially for a {\em forward} problem: given the interactions in a system (as well as external fields etc.\ where relevant), predict the dynamics of the order parameters of interest. In recent years, there has been strong interest in the {\em inverse} problem: given observations of the order parameter dynamics, find the interactions; for a review see e.g.\ \cite{roudi2015multi}. This interest has been generated by recent advances in recording technology in various fields of science, allowing massive data sets to be gathered that invite researchers to attempt to reverse-engineer the underlying systems from the observed data. 

The path integral formulation for forward problems described here is a natural first step in finding inverse equations, which in the simplest case consist of an inversion of the forward equations. Furthermore, in some cases, the path integral method can be immediately applied to the inverse problem itself. For instance, in the presence of hidden degrees of freedom, when trying to reconstruct the interactions in a spin model by observing only some of the spins and not the others, then calculating the likelihood of the data involves tracing over the trajectory of the hidden spins. This can then be done using the path integral methods discussed in this review \cite{dunn2013learning}.

{\em The extended Plefka expansion}.
In a spin system at equilibrium, the  magnetization of each spin is really the only parameter of interest, with spin-spin correlations then determined indirectly from the linear response of the system. This is not the case for out-of-equilibrium dynamics. Here out-of-equilibrium correlation and response functions need to be calculated in addition to the magnetizations in order to achieve a full statistical description of the system. 
So while at equilibrium and also in the simplest form of the dynamical TAP approach for Ising models~\cite{roudi2011dynamical} it makes sense to perform a Plefka expansion by fixing only the magnetization (and in the dynamics, the conjugate fields), in the full dynamical treatment response and correlation functions should also be fixed as order parameters. This is even more important for soft spin models, where even at equilibrium one would want to use at least the means and variances of the local degrees of freedom as order parameters.
Such ``extended" Plefka expansions have been used for the so-called $p$-spin spherical spin glass model in \cite{biroli1999dynamical} and, more recently, for general systems of coupled Langevin equations \cite{bravi2015extended} and Ising spin glasses \cite{Bachschmid-Romano2016extended}. They should be a productive route for further progress in the field, e.g.\ by extending them further to incorporate inference from observed data~\cite{Bravi_inference_arxiv}.

\ack
%\section*{Acknowledgement}
This work has been partially supported by the Marie Curie Initial Training Network NETADIS (FP7, grant 290038). YR was also financially supported by the Kavli Foundation, Norwegian Research Council  (grant number 223262) and the Starr Foundation. The authors are also grateful to Luca Peliti for fruitful discussions. 

\appendix
\section*{Appendix}
\setcounter{section}{1} 

The Legendre transform approach starts from the saddle point value of the log generating functional
\bea
\ln Z_s[\psi]&=& \sum_t i \hat{\bm h}^s(t)\cdot[{\bm h}^s(t)-\bm h^{\rm ext}] \label{eq:logZs}\\
&&{}+\sum_{it} \left \{\lc\big(\psi_i(t+1)+h^s_i(t)-i(\bm J^{\rm T}\hat{\bm h}^s(t+1))_i\big) -\lc(h^s_i(t))\right \}. \nonumber
\eea
We first rewrite this expression in a form that will simplify some of the algebra below, by decomposing the function $\lc(\cdot)$ in \eref{eq:logZs} as
\bea
\lc(x) &=& H_2[\tanh(x)] + x \tanh(x)\\
H_2[x]&\equiv& -\frac{1+x}{2} \ln \frac{1+x}{2}-\frac{1-x}{2} \ln \frac{1-x}{2}.
\eea
Here $H_2[x]$ is simply the entropy of a single spin with magnetization $x$. This decomposition gives, bearing in mind the definition of $\mu_i(t)$ in\eq{eq:mu_def},
\bea
\fl \ln Z_s [\psi]&=&\sum_{it} H_2[\mu_i(t)] + \sum_t i \hat{\bm h}^s(t)\cdot[{\bm h}^s(t)-\bm h^{\rm ext}] \label{eq:logZs2}\\
\fl &&{}+\sum_{t}\bm \mu(t+1)\cdot[\bm \psi(t+1)+\bm h^s(t)]-i\sum_k \hat{h}^s_k(t) \sum_i \mu_i(t) J_{ki}
%\nonumber \\
%&&{}
-\sum_{it} \lc(h^s_i(t)). \nonumber
\eea
The second sum in the second line
% of \eref{eq:logZs2} 
now cancels the second sum in the first line because of the saddle point equation\eq{eq:sp2}, and we end up with the expression
\begin{eqnarray}
\ln Z_s[\psi]= \sum_{i,t} H_2[\mu_i(t)]+\sum_{t} \bm \mu(t+1)\cdot [\bm \psi(t+1)+\bm h^s(t)]-\sum_{it} \lc(h^s_i(t)).
\label{eq:logZs3}
\end{eqnarray}
The Legendre transform of this expression with respect to the magnetizations $m_i(t)$ is
\be 
\Gamma_s[m]=\ln Z_s[\bm \psi]-\sum_{t} \bm \psi(t)\cdot \bm m(t),
\ee
where the $\psi$ are to be expressed in terms of the $m$ by solving
\begin{equation}
m_i(t)=\frac{\partial \ln Z_s[\psi]}{\partial \psi_i(t)}
\end{equation}
We already know that $\partial \ln Z_s[\psi]/\partial \psi_i(t)=m_i^s(t)=\mu_i(t)$, so at this saddle point level one has $m_i(t)= \mu_i(t)$. The second saddle point equation\eq{eq:sp2} then becomes $\bm h^s(t) = \bm h^{\rm ext}+\bm J \bm m(t)$ and inserting this gives 
\be
\fl 
\Gamma_s[m]=\sum_{it} H_2[m_i(t)]+\sum_{it} m_i(t+1) \Big[h^{\rm ext}_i+\sum_jJ_{ij} m_j(t)\Big]-\sum_{it} \lc\Big(
h^{\rm ext}_i+\sum_jJ_{ij} m_j(t)
\Big).
\label{eq:Gamma_s}
\ee
The equation of state is, by standard Legendre transform properties,
\begin{equation}
-\psi_i(t)=\frac{\partial \Gamma_s[m]}{\partial m_i(t)}\ ,
\end{equation}
which yields
\begin{equation}
\fl 
-\psi_i(t)=
%
%\frac{\partial \Gamma_s}{\partial m_i(t)}=%
-\tanh^{-1}(m_i(t))+h^{\rm ext}_i+\sum_{j} J_{ij} m_j(t-1)+\sum_j J_{ji} m_j(t+1)-\sum_j J_{ji} \tanh(h^s_j(t)).
\end{equation}
At $\psi_i(t)=0$ this is naturally solved by $m_i(t+1)=\tanh(h^{\rm ext}_i+\sum_j J_{ij} m_j(t))$. So far we have not achieved anything new: we have merely provided an alternative way of obtaining the naive mean-field equations\eqq{eq:naive_mf1}{eq:naive_mf2} we already had.

Now let us consider the Legendre transform of the generating functional including the Gaussian correction, $\ln Z[\psi]=\ln Z_s[\psi]+ A[\psi]$
\be
\Gamma[m] = \ln Z_s[\psi] + A[\psi] -\sum_{t} \bm\psi(t)\cdot \bm m(t),
\label{eq:Gammamu}
\ee
where as before $\psi$ is to be treated as a function of the magnetizations, as determined by the condition
%
%As we saw in the previous section, we expect the magnetizations to be modified to
\begin{equation}
m_i(t)=\frac{\partial \ln Z[\psi]}{\partial \psi_i(t)}=\mu_i(t)+A_i(t).
\end{equation}
Here we have used the earlier definition $A_i(t)=\partial A/\partial\psi_i(t)$.

Expressing the right hand side of \eref{eq:Gammamu} in terms of $m$, and keeping only terms linear $A_i$, we obtain, as shown below, 
\begin{eqnarray}
\Gamma[m]&=& \sum_{it} H_2[m_i(t)]+\sum_{it} m_i(t+1) \Big[h^{\rm ext}_i +\sum_{j} J_{ij} m_j(t)\Big]
\\
&&{}-\sum_{it} \lc\Big (h^{\rm ext}+\sum_{j} J_{ij} m_j(t)\Big )+A[m].
\nonumber
\label{eq:tap-like}
\end{eqnarray}
Remarkably, all the $A_i$ terms have cancelled here and the only difference to the saddle point result $\Gamma_s[m]$ in\eq{eq:Gamma_s} is the naive addition of the correction term $A$, expressed in terms of $m$.

From $\Gamma[m]$ we can now derive the equation of state for the magnetizations from
\begin{eqnarray}
\fl -\psi_i(t)=\frac{\partial \Gamma[m]}{\partial m_i(t)} &=& - \tanh^{-1}(m_i(t)) +h^{\rm ext}_i+ \sum_j J_{ij} m_j(t-1) +\frac{\partial A}{\partial m_i(t)} \\
&&{}+ \sum_k J_{ki} \Big[m_k(t+1) -\tanh\big(h^{\rm ext}_k+\sum_j J_{kj} m_j(t)\big)\Big].
\end{eqnarray} 
If we are only interested in corrections to {\em quadratic} order in $J$ for the field acting on each spin, we can drop the terms in the second line of the equation above as the term in square brackets is already $\order(J^2)$. In the physical limit $\psi \to 0$ one then obtains
\be
m_i(t)= \tanh\Big(h^{\rm ext}_i+ \sum_j J_{ij} m_j(t-1) +
%\lim_{\psi \to 0}
\frac{\partial A}{\partial m_i(t)}  \Big). \label{eq:mLGcorr}
\ee
We can now appreciate the role played by the Legendre transform: the corrections we obtain act on the effective fields and so are inside the $\tanh$. As explained in the main text this makes more physical sense. 

To evaluate $\partial A/\partial m_i(t)$, one can start from the expression\eq{eq:A_linearized} for $A$. Replacing $h^s_i(t)$ in this using the saddle point equation\eq{eq:sp2} gives
\bea
A &=& -\frac{1}{2}\sum_{it} \left[\mu^2_{i}(t+1)-\tanh^2\Big(h^{\rm ext}_i+\sum_j J_{ij} \mu_i(t)\Big)
\right] \sum_k J^2_{ik} [1-\mu^2_k(t)].
\eea
As $A$ is already $\order(J^2)$, replacing all $\mu_i(t)$ by $m_i(t)$ in this expression only gives a negligible correction of $\order(J^4)$. Also the first square bracket is small at the physical saddle point, of $\order(J^2)$, so the derivative of the final factor with respect to $m_i(t)$ can be dropped. Finally the derivative of the tanh can also be neglected as it is of order $J$. One thus finds to leading order the simple result
\be
\frac{\partial A}{\partial m_i} = - m_i(t) \sum_{k} J^2_{ik} (1-m^2_k(t-1))+\order(J^3),
\ee
Combined with \eref{eq:mLGcorr} this yields the dynamical TAP equations \eref{eq:DynTAP}.

It remains to show\eq{eq:tap-like}.
Using the expression \eref{eq:logZs3} for $\ln Z_s$ and $m_i(t)=\mu_i(t)+A_i(t)$ in \eref{eq:Gammamu}, we can write $\Gamma$ as 
\be
\fl \Gamma[\bm m]= \sum_{i,t} H_2[\mu_i(t)]+\sum_{t} \bm \mu(t) \cdot \bm h^s(t-1)-\sum_{it} \lc(h^s_i(t))-\sum_i \psi_i(t) A_i(t) +A. \label{eq:GammamuApp}
\ee
We have shifted the time index $t$ by one in the second and fourth sum for later convenience. 
Now we express the right hand side of \eref{eq:GammamuApp} in terms of $m$ and keep only terms linear in $A_i$. For the last sum we need 
the following identity, which can be obtained from the
definition of $\mu_i(t)$ in \eref{eq:mu_def} 
together with $m_i(t)=\mu_i(t)+A_i(t)$:
\be
\fl \psi_i(t) = \tanh^{-1}(m_i(t)-A_i(t)) -h^{\rm ext}_i- \sum_j J_{ij} m_j(t-1) +\sum_j J_{ij} A_j(t-1)+i(\bm J^{\rm T} \hat{{\bm h}}^s(t))_i.
\ee
To linear order in $A_i$, the various terms on the right hand side of \eref{eq:GammamuApp} are then
\bea
\fl H_2[\mu_i(t)] &=& H_2[m_i(t)]+ A_i(t) \tanh^{-1}(m_i(t)) \\
\fl \mu_i(t) h^s_i(t-1) &=& m_i(t)\Big [h^{\rm ext}_i+\sum_j J_{ij} m_j(t-1)\Big]-A_i(t) \sum_{j} J_{ij} m_j(t-1)\nonumber \\
&&{} -m_i(t)\sum_{j}  J_{ij} A_j(t-1) 
-h^{\rm ext}_i A_i(t) 
\\
\fl -\lc(h^s_i(t)) &=& - \lc\Big(h^{\rm ext}_i+\sum_j J_{ij}m_j(t)\Big) + \tanh\Big(h^{\rm ext}_i+\sum_k J_{ik}m_k(t)\Big) \Big[\sum_j J_{ij} A_j(t)\Big] \\
\fl -\psi_i(t) A_i(t) &=&  -A_i(t) \tanh^{-1}(m_i(t))+A_i(t)\sum_j J_{ij} m_j(t-1)-iA_i(t)\sum_j J_{ji} \hat{h}^s_j(t)\\
\nonumber
&&{}+h^{\rm ext}_i A_i(t).
\eea
Putting all these together we notice that the second term on the right hand side of the first equation above and the first term of the last equation, the second terms in the second and last equations, as well as the last terms in the second and last equations cancel each other, yielding 
\bea
\fl \Gamma[m]&=& \sum_{it} H_2[m_i(t)]+\sum_{it} m_i(t) \Big[h^{\rm ext}_i+\sum_{j}J_{ij} m_j(t-1)\Big]-\sum_{it}  \lc\Big(h^{\rm ext}_i+\sum_j J_{ij}m_j(t)\Big)
\nonumber\\
\fl&&{}-\sum_{ijt}  J_{ij} A_j(t-1)m_i(t)+\sum_{it} \tanh\Big(h^{\rm ext}_i+\sum_k J_{ik}m_k(t)\Big) \Big[\sum_j J_{ij} A_j(t)\Big] \nonumber\\
\fl &&{}-i\sum_{ijt} A_j(t) J_{ij} \hat{h}^s_i(t)+A+\order(A^2_i).
\label{eq:GammaG2}
\eea
Using the fact that from the first saddle point equation\eq{eq:sp1}
\be
-i\hat{h}^s_i(t) = \mu_i(t+1) -\tanh(h^s_i(t)) =  m_i(t+1) - \tanh(h^s_i(t)) - A_i(t+1),
\ee
we can write the terms in the second line of \eref{eq:GammaG2}, together with the first term in the third line, as
\be
\sum_{ijt} \Big\{-\tanh(h_i^s(t))+\tanh\Big (h^{\rm ext}_i+\sum_k J_{ik}m_k(t)\Big)\Big\} J_{ij} A_j(t) +\order(A_i^2) 
\ee
As the term in curly braces is itself a correction term that is nonzero only because of the difference between $\mu_i(t)$ and $m_i(t)$, this overall expression can be
neglected as subleading. The remaining terms of\eq{eq:GammaG2} then give exactly
\eref{eq:tap-like} as claimed. 
Note that keeping only terms linear in $A_i$ in this calculation -- which then eventually cancel -- in some sense plays the same role as expanding to second order in $J^2$ in the Plefka method, as $A_i$ is of $\order(J^2)$.
\end{fmffile}
\section*{References}
\bibliography{mybibliographydynamics}
\end{document}